\documentclass[a4paper,aps,prb,reprint,superscriptaddress,floatfix]{revtex4-2}
\usepackage{graphicx}
\usepackage{amsmath,amssymb}
\usepackage{epstopdf}
\usepackage{xcolor}
\usepackage[english]{babel}
\usepackage[utf8]{inputenc}
\usepackage{bm,siunitx}
\usepackage{booktabs}
\newcommand{\be}{\begin{equation}}
	\newcommand{\ee}{\end{equation}}

\usepackage{hyperref}
\hypersetup{colorlinks,allcolors=red}

\begin{document}
	
	\title{Interpretation of spin-wave modes in Co/Ag nanodot arrays probed by broadband ferromagnetic resonance}
	
	\author{Daniel Mark\'o}
	\affiliation{Universit\'e Paris-Saclay, UVSQ, CNRS, GEMaC, 78000 Versailles, France}
	\affiliation{Silicon Austria Labs GmbH, Magnetic Microsystem Technologies, Europastr.~12, 9524 Villach, Austria}
	
	\author{Rajgowrav Cheenikundil}
	\affiliation{Universit\'e de Strasbourg, CNRS, Institut de Physique et Chimie des Mat\'eriaux de Strasbourg, F-67000 Strasbourg, France}
	
	\author{Julien Bauer}
	\affiliation{Universit\'e de Strasbourg, CNRS, Institut de Physique et Chimie des Mat\'eriaux de Strasbourg, F-67000 Strasbourg, France}
	
	\author{Kilian Lenz}
	\affiliation{Institute of Ion Beam Physics and Materials Research, Helmholtz-Zentrum Dresden-Rossendorf, 01328 Dresden, Germany}
	
	\author{Wan-Chen Chuang}
	\affiliation{Department of Materials Science and Engineering, National Chung Hsing University, Taichung 402, Taiwan}
	
	\author{Ko-Wei Lin}
	\affiliation{Department of Materials Science and Engineering, National Chung Hsing University, Taichung 402, Taiwan}
	
	\author{Jong-Ching Wu}
	\affiliation{Department of Physics, National Changhua University of Education, Changhua 500, Taiwan}
	
	\author{Massimiliano d'Aquino}
	\affiliation{Department of Electrical Engineering and ICT, University of Naples Federico II, Naples, Italy}
	
	\author{Riccardo Hertel}
	\affiliation{Universit\'e de Strasbourg, CNRS, Institut de Physique et Chimie des Mat\'eriaux de Strasbourg, F-67000 Strasbourg, France}
	
	\author{David S.~Schmool}
	\email{Corresponding author: david.schmool@uvsq.fr}
	\affiliation{Universit\'e Paris-Saclay, UVSQ, CNRS, GEMaC, 78000 Versailles, France}
	
	\date{\today}
	
\begin{abstract}

We present a detailed investigation of the magnetization dynamics in Co/Ag nanodots, which due to their size can support standing spin-wave (SSW) modes with complex spectral responses. To interpret the experimentally measured broadband vector network analyzer ferromagnetic resonance data, we compare the spectra of the nanoarray structure with those of the unpatterned Co/Ag film of identical thickness, which serves as a baseline for obtaining the general magnetic parameters of the system. Using a novel frequency domain, matrix-free simulation method of the dynamic response, we identify the nature of the excitation modes, which allows us to assess the boundary conditions for the nanodots. We find an excellent agreement between the calculated and experimental values for the frequencies of the fundamental (uniform-like) (011) mode. The existence of an edge-localized mode in the experiment has been confirmed and fits very well with theory and micromagnetic simulations, having the form of a flapping mode at the extrema of the nanodot in one of the in-plane directions. Its frequency is below the fundamental mode’s frequency and has been shown to be a consequence of the imaginary wave vector for such localized SSW modes. Higher order SSW modes can be generated from the theory, which allows us to find a probable mode number for the second bulk SSW (201 or 221 or 131), which lies at frequencies above the fundamental mode.
\end{abstract}

	
	\maketitle
	
\section{Introduction}
	
The study of the magnetic properties of nanostructured arrays or nanodots, is a key area of research in modern magnetics and a central field in nanomagnetism \cite{Guimaraes2017,Binns2014,Schmool2021a}. These systems display a number of properties, which are modified due to the spatial confinement of the magnetic layer as well as surface anisotropy effects. This confinement includes both the vertical thickness and the lateral dimension of the nanodots. 
	
Of the various modifications of the magnetic properties of magnetic nanodots arrays, with respect to the bulk properties of the constituent materials, magnetization, $g$-factor, and magnetic anisotropies can be altered. Furthermore, dipolar coupling between the dots can also be important, though it can be controlled and reduced by using a sufficiently large array period. Such considerations are central to the understanding of the dynamic magnetic properties of these systems, as measured using ferromagnetic resonance (FMR) \cite{Schmool2007}. 
	
FMR provides a versatile and sensitive probe of the magnetic state of low dimensional systems, such as nanostructured arrays, and is an ideal tool to study the magnetization dynamics \cite{Heinrich1994,Schmool2009,Farle1998}. Furthermore, since the resonance is sensitive to the local {effective} internal field, any interactions via dipolar or exchange coupling can also be detected and studied by FMR \cite{Lindner2003,Ovsyannikov2019,Zivieri2000,Demokritov1998,Layadi2002,Layadi2004,Schmool2009a,Schmool2007a}. Of particular importance is its sensitivity to the magnetic boundary conditions that determine the allowed wave vectors in standing spin-wave modes. These are responsible for the mode patterns, and hence, resonance frequency--field [$f(H)$] characteristics of the magnetic system \cite{Rado1959,Puszkarski1979,Puszkarski2017}. The excitation modes in low-dimensional systems are intimately related to the specific magnetic properties of a sample and any changes will alter the $f(H)$ characteristics, making FMR an extremely sensitive tool \cite{Maksymowicz1986,Schmool1998}. 
	
In this paper, we describe the detailed experimental study of Ag-capped Co nanodots, which are expected to have weak interactions due to the spatial separation between the structures. We outline the procedure for the analysis of the experimental data obtained from the broadband FMR measurements and its subsequent interpretation. Indeed, this is not a simple process due to a number of technical issues, which we will discuss in the paper. We provide an outline of the theoretical framework necessary for a correct understanding of the resonance absorption phenomena observed. In conjunction with this theory, we employ simulations based on a recently-developed frequency-domain micromagnetic approach \cite{d’Aquino2023}, which is able to effectively treat arbitrarily-shaped systems (e.g., those studied in Ref.~\cite{cheenikundil_high-frequency_2022}) and assists in the designation of the expected excitation modes and ultimately provides a working model to evaluate the wave vectors of the excited modes for the circular nano\-structures used in this study. Using the film system, based on the layered structures employed, allows us to perform reference FMR measurements to obtain the material constants for the Co nanostructures. These are subsequently used in the calculations of the expected excitation modes of the system. 
	
It should be noted that a direct interpretation of such experiments on nanosystems can be compounded by the consideration of the three-dimensional boundary conditions, and this can only be solved using a model for the pinning conditions in all spatial directions. We overcome this difficulty by using micromagnetic simulations to provide a likely mode pattern, from which we assess the boundary condition. We provide a full description of this analysis, an in-depth overview of the calculations used, and a comparison to the experiment. 
	
It is important to note that the analytical model represents a crude approximation for nanostructures, where edge and confinement effects compound the situation and standing spin wave modes cannot be correctly represented as plane waves, such as in the case of extended film geometries. To assist the analysis for the approximation of the wave vectors, we will use micromagnetic simulations. These provide a much better estimate of the spatial distributions for the standing spin wave modal patterns and can be used as an indicator of the boundary conditions, which can be applied to the analytical model. While, as we mentioned, this method is approximative, it does provide a useful insight and a reasonable estimate for a comparison of the analysis methods. 

\section{The Theory of Ferromagnetic Resonance in Magnetic Nanodot Structures} 
The theory of ferromagnetic resonance is well documented \cite{Smit1955,Vonsovskii1966,Yalcin2013} and we can use a general form of the resonance equation, which is derived from the Landau-Lifshitz equation \cite{Landau1935,Gilbert1955}. This approach allows us to use the relevant terms of the free energy contributions relevant to the sample in question. These will specifically include the magnetic anisotropies, which are related to the sample structure for the magnetocrystalline anisotropies and the sample geometry for the shape contribution. In addition, we can include effects such are coupling and exchange by adding these specific terms into the free energy. For low dimensional systems, such as thin films and nanostructures, boundary effects can be taken into account by imposing the pinning conditions, which account for the modification of the spin freedom at the edges of the magnetic entity. This will affect the wave vectors of standing spin-wave modes \cite{Puszkarski1979}. It is worth noting that the application of a time-varying magnetic field, such that the total applied field is given by $\mathbf{H} (\mathbf{r}, t) = \mathbf{H}_0 +\mathbf{h}(\mathbf{r}, t)$, where  $\mathbf{H}_0$ is the static component and  $\mathbf{h}$ the dynamic component, implies corresponding static and dynamic components of the magnetization, $\mathbf{M}(\mathbf{r}, t)  = \mathbf{M}_0 +\mathbf{m}(\mathbf{r}, t)$. In the absence of magnetic anisotropies, the static external field will cause the equilibrium magnetization to align along the same direction. The excitation field, $\mathbf{h}(\mathbf{r}, t)$, under the condition $||\mathbf{h}|| \ll H_0$, will produce wave-like perturbations in the magnetization called spin waves, as described by the spatial and temporal variation of $\mathbf{m}(\mathbf{r}, t) \sim \mathbf{m}(\mathbf{r}) e^{i\omega t}$. 

\subsection{Resonance equations}
\label{FMRnano}
Using the contribution of the spatial variation of the magnetization to the time derivative of the magnetization $\partial \mathbf{M}/\partial t$, the dipole-exchange spin-wave dispersion relation for an infinite ferromagnetic medium can be written in the general form as \cite{Herring1951,Vanderveken2021}:
\begin{equation}
	f^2\!=\!\left(\!\frac{\mu_0 \gamma}{2\pi}\right)^2\!(H\!+\!H_K\!+\!Dk^2)(H\!+\!H_K\!+\! Dk^2\!+\!M\sin^2\theta_\textrm{k}).
	\label{eq:SWresInf}
\end{equation}

Here, $H$ is the static applied magnetic field, $H_K$ is the anisotropy field to be defined with respect to the crystalline axes and the orientation of the applied field, $\theta_\textrm{k}$ defines the angle between the directions of the wave vector and the static magnetization,$\ D = 2 A_\text{ex}/\mu_0 M_s$ is the spin wave constant, which depends on the exchange stiffness constant, $A_\text{ex}$, and is related to the exchange constant and the crystalline structure of the magnetic sample, while $k$ denotes the wave vector of spin wave mode excitations. Equation (\ref{eq:SWresInf}), known as the Herring-Kittel equation, is one of the fundamental expressions of the resonance condition for a ferromagnetic sample and must be adapted to the specific experimental conditions. In this formulation, the static component of the magnetization, $\mathbf{M}$, is taken to be in the direction perpendicular to the dynamic component of the magnetization.  Prior knowledge of the magnetic properties of the sample is desirable, but the fitting of the resonance curve to experimental data is a good method to determine magnetic properties of samples (saturation magnetization, anisotropy constants and the $g$-factor). When using this formulation, it is useful to bear in mind the limits of frequency, which will depend on the nature of the spin-wave excitations. In the long wavelength limit ($Dk^2 \ll 1$), the spin-wave terms will vanish and $f \simeq (\mu_0 \gamma/2\pi) \sqrt{(H+H_K)(H + H_K+ M\sin^2\theta_\textrm{k})}$. This dispersion relation is valid for dipolar spin waves. For $\theta_\textrm{k} =0$, the dispersion relation reduces to $f = (\mu_0 \gamma/2\pi)[H + H_K]$, while for $\theta_\textrm{k} = \pi/2$ we have  $f = (\mu_0 \gamma/2\pi)\sqrt{(H+H_K)(H + H_K+ M)}$. At the other end of the spin-wave scale, for short wavelength spin waves, the spin wave term will dominate the dispersion relation, giving $f \simeq (\mu_0 \gamma/2\pi) Dk^2$. This limit is valid when the exchange interaction is dominant. From the general dispersion relation, we can note that, while the exchange spin waves are isotropic, dipolar spin waves are anisotropic due to the angle $\theta_\textrm{k}$. 
	
In the above, the properties of spin waves in an infinite bulk medium were considered. We will now turn our attention to the case of spin waves in ferromagnetic thin films, where the in plane dimensions are infinite and the film thickness will define the boundaries of the ferromagnetic medium. In this approximation we consider that the film thickness, $L$, is sufficiently small with respect to the ferromagnetic skin-depth. For a typical experiment in the GHz regime, the skin-depth will be in the micron range, so for thin films with thickness in the tens of nm, this approximation is valid. Once again, the components of the dynamic magnetization will form a plane wave (spin wave). While the film boundaries will not affect the exchange field, they will introduce magnetic boundaries and will affect the dipolar field. The dispersion relation for spin waves can be shown to take the form \cite{Kalinikos1986,Vanderveken2021}: 
\begin{equation}
		\omega_n^2 = (\omega_H +  \omega_D k_n^2)[\omega_H +  \omega_D k_n^2 +  \omega_M F_{nn}(k_{\zeta}L) ],
		\label{eq:DispGen}
\end{equation}
where $\omega_H = \gamma \mu_0 H$, $\omega_D = \gamma \mu_0 D$ and $\omega_M = \gamma \mu_0 M$. The function$\ F_{nn}(k_{\zeta}L)$ is the matrix element of the magnetic dipole interaction and $\ n = 0, 1, 2, ...$  is a quantization number for so-called perpendicular standing spin waves (PSSW), which can be expressed in the form:
\begin{widetext}	
\begin{equation}
		F_{nn}(k_{\zeta}L) = P_{nn} + \sin^2 \theta \left[1 - P_{nn} (1+ \cos^2 \phi) + \omega_M \frac{P_{nn} (1-P_{nn} )\sin^2 \phi}{\omega_H + \omega_D k_n^2} \right]
		\label{eq:Fnn}
\end{equation}
\end{widetext}
with $\ k_n^2 = k_{\zeta}^2 + \kappa_n^2$, where $\kappa_n$ is the transverse wave vector. We note that the matrix element, $P_{nn}$, lies in the range $0\leq P_{nn} \leq 1$ for $0 \leq k_{\zeta} L \leq \infty$. In the long wavelength limit ($k_{\zeta} L \ll 1$), simple expressions can be obtained for $P_{nn}$. For example, in the case of perfect pinning \cite{Kalinikos1986}:
\begin{equation}
		P_{nn'} = \frac{k^2_{\zeta}}{k^2_{n'}} \delta_{nn'} +\frac{k^2_{\zeta}}{k^2_{n}} \frac{\kappa_n \kappa_{n'}}{k^2_{n'}} F_n\left(\frac{1 + (-1)^{n +n'}}{2}\right),
		\label{eq:PnnPP}
\end{equation}	
while for unpinned boundaries this becomes:
\begin{widetext}
\begin{equation}
		P_{nn'}\!=\!\frac{k^2_{\zeta}}{k^2_{n'}}\delta_{nn'}\!-\!\frac{k^4_{\zeta}}{k^2_{n} k^2_{n'}}\!F_n\!\left(\!\frac{1+ (-1)^{n +n'}}{2[(1+\delta_{0n})(1+\delta_{0n'})]^{1/2}}\!\right).
		\label{eq:PnnUP}
\end{equation}
\end{widetext}	
For this latter simple case, in a thin film, we obtain $P_{nn'} = \frac{k_{\zeta} L}{2}$ for $n= 0$ and $P_{nn'} = \left(\frac{k_{\zeta} L}{n\pi}\right)^2$for $n\neq 0$. For a spin wave propagating in the plane of the magnetic film and perpendicular to the applied bias field, $\ k_x = 0, k_y = k_{||}$, we can write:
\begin{equation}
		F_{nn}(k_{\zeta}L) =  1+ \frac{ \omega_M }{\omega_H + \omega_D k_n^2} P_{nn} (1-P_{nn}).
		\label{eq:Fnn2}
\end{equation}
For the lowest body mode,$\ (n = 0)$, $\ P_{nn}$ can be expressed as:
\begin{equation}
		P_{00} =  1- \frac{ 1 - e^{-k_{||}L}}{k_{||}L}. 
		\label{eq:P00}
\end{equation}
For more general expressions, see Ref.~\cite{Kalinikos1986}. 
	
In the case of out-of-plane magnetized circular disks, Dobrovolskiy \textit{et al.} \cite{Dobrovolskiy2020} and Kakazei \textit{et al.} \cite{Kakazei2004} considered the excited spin-wave eigenmodes as being described by Bessel functions of the zeroth order due to the axial symmetry of the structure. The dynamic in-plane magnetization,$m_x, m_y \propto J_0 (k \rho)$, where $\rho$ is the radial coordinate. Assuming that the static disk magnetization does not depend on the thickness coordinate z, one can use a spin-wave dispersion equation similar to the one used for infinite films. The resonance equation takes the form \cite{Dobrovolskiy2020}:
	\begin{equation}
		\omega_n^2 = (\omega_H +  \omega_D k_n^2)(\omega_H +  \omega_D k_n^2 +  \omega_M F_{nn}(\kappa_n \beta) ),
		\label{eq:DispGen2}
	\end{equation}
	where we have quantized radial wave vector$\ k_n = \kappa_n/R$ and the effective magnetic field becomes:$\ H_\text{eff} = H + H_{nn}$. This expression is essentially equivalent to Eq.~(\ref{eq:DispGen}). The second term in this magnetic field corresponds to the diagonal matrix elements of the static dipolar field, given as:
	\begin{widetext}
	\begin{equation}
		H_{nn} = \mu_0 M \left( \int_0^{\infty} f(\beta x) J_1 (x) dx \int_0^1 \rho \phi_n^2(\rho) J_0 (\rho x) d\rho -1\right),
		\label{eq:DipMat}
	\end{equation}
	\end{widetext}
	where $\beta = L/R$ is the thickness-radius aspect ratio. In the limit of small $\beta$ and strong dipolar pinning of the dynamical component of the magnetization at the edges of the nanodot, the value of $\kappa_n$ becomes the$\ n^{th}$ root of $\ J_0 (\kappa) = 0$ with $n$ being an integer value. 
	\begin{equation}
		\phi_n(\rho) = \frac{1}{\sqrt{N_n}} J_0 (\kappa_n \rho)
		\label{eq:Phi}
	\end{equation}
	and
	\begin{equation}
		N_n = \frac{1}{2} [J_0^2 (\kappa_n) + J_1^2 (\kappa_n) ]
		\label{eq:Norm}
	\end{equation}
	is the normalization factor for $\phi_n (\rho)$. For the special case of an unlimited medium, Eq.~(\ref{eq:DispGen2}) takes the form of the Herring-Kittel equation as given in Eq.~(\ref{eq:SWresInf}). 
	
	In the above, the edge spins are perfectly pinned and give rise to circular modes as described in Ref.~\cite{Dobrovolskiy2020}. Unpinning at the edge boundary can also be considered and should give rise to edge localized \textit{flapping}-like modes. Radial modes can also be excited, giving, in theory, rich spin-wave spectra. The specific number of observable spin-wave modes will depend explicitly on the lateral and thickness dimensions as well as on the exchange-stiffness constant of the nanodots.

	\subsection{Spin-wave vectors}
	As can be seen from the previous section, to obtain the correct form of the dispersion relation for dipole-exchange spin waves, we require the wave vector for the spin-wave excitations. These can be calculated from the Rado-Weertman boundary equations, which can be expressed in the form \cite{Rado1959}:
	\begin{align}
		\frac{\partial m_x (\zeta)}{\partial \zeta} + \delta \cos 2 \theta ~m_x (\zeta) &= 0,
		\label{eq:RW1}\\
		\frac{\partial m_y (\zeta) }{\partial \zeta} + \delta \cos ^2 \theta ~m_y (\zeta) &= 0.
		\label{eq:RW2}
	\end{align}
	Here, $m_{x, y} (\zeta, t) \propto m_{x, y} e^{i(\omega t - k_{\zeta} \zeta)}$ denotes the transverse components of the magnetization and $k_{\zeta}$ indicates the longitudinal component of the wave vector. Analysis of the boundary conditions and pinning parameters can then enable the evaluation of the allowed wave vectors for the spin-wave excitations \cite{Maksymowicz1986,Schmool1998}:
	\begin{equation}
		[\delta_1^p \delta_2^p - (k_n^p)^2] \tan (k_n^p L) = k_n^p L ( \delta_1^p + \delta_2^p).
		\label{eq:KV}
	\end{equation}
	Here, the pinning parameters for the transverse moments are expressed as:
	\begin{align}
		\delta_{1, 2}^x &= \delta_{1, 2}\cos 2 \theta
		\label{eq:PP1}\\
		\delta_{1, 2}^y &= \delta_{1, 2}\cos^2 \theta.
		\label{eq:PP2}
	\end{align}
	The uniform ferromagnetic resonance mode has a value of $\ k = 0$, where all spins in the system are considered to precess in phase throughout the sample, while non-zero $k$ correspond to the spatial variation of the transverse components of the magnetization. It is worth noting, that, depending on the specific boundary conditions, the uniform mode is not always excited and does not necessarily correspond to the fundamental mode in the spectra of low-dimensional systems. In the following, we will outline how we can take into account the boundary conditions to evaluate the wave vectors in increasingly complex situations. In the simplest case of a single magnetic layer, the wave vectors can be considered in one dimension and expressed in a simple form:	
	\begin{equation}
		k_\text{pp} = \frac{p \pi}{L},
		\label{eq:WV1Dpp}
	\end{equation}
	
	\noindent where$\ L$ defines the layer thickness. This expression considers that the surface spins are perfectly pinned, i.e., the surface or boundary spins are completely fixed and do not precess. In this case, the fundamental mode will not correspond to the uniform FMR precession, as noted above. Here we obtain the ideal Kittel expression for the resonance equation. We note that this can be used to generate the spin-wave spectrum of standing wave modes for integer $p$ values. We can extend this model to consider the case of free pinning, where the surface spins have bulk freedom of precession, for which we can write
	\begin{equation}
		k_\text{fp} = \frac{(p - 1) \pi}{L}.
		\label{eq:WV1Dpf}
	\end{equation}
	In this particular case, the fundamental mode is indeed the uniform FMR excitation. Taking the argument further, we can also generalize for intermediate pinning, in which the surface pinning parameter, $\delta$, lies between these two limits (perfect freedom and perfect pinning). In this case, we can express the wave vector as:
	\begin{equation}
		k_\text{ip} = \frac{(p - \delta) \pi}{L},
		\label{eq:WV1Dip}
	\end{equation}
	where we take $\ 0 \leq \delta \leq 1$. In Eqs.~(\ref{eq:WV1Dpp}), (\ref{eq:WV1Dpf}) and (\ref{eq:WV1Dip}), we take the pinning to be symmetric, i.e., both boundaries having identical pinning conditions. For asymmetric pinning, we must write the pinning parameters for the two interfaces explicitly as: 	
	\begin{equation}
		k_{A} = \left[ p - \frac{(\delta_1 + \delta_2) }{2} \right] \frac{\pi}{L},
		\label{eq:WV1Das}
	\end{equation}
	with the two pinning parameters, $\delta_1$ and $\delta_2$, taking values between 0 and 1 for the two extreme cases of pinning considered above. 
	
	
	
	We can extend the above arguments to define a three-dimensional structure with boundaries and hence specific pinning parameters in each of the spatial directions:
\begin{widetext}
	\begin{equation}
		k_{3D-A}^2 = \left[\left(m - \frac{(\delta_{x1} + \delta_{x2}) }{2}\right)\frac{\pi}{L_x}\right]^2 + \left[\left(n - \frac{(\delta_{y1} + \delta_{y2})}{2} \right) \frac{\pi}{L_y}\right]^2 
		+\left[\left(p - \frac{(\delta_{z1} + \delta_{z2})}{2}\right)  \frac{\pi}{L_z}\right]^2. 
		\label{eq:WV3Dasym}
	\end{equation}
\end{widetext}

	Here, we have the modal numbers$\ m, n$ and$\ p$ for the three orientations of the nanostructure with dimensions $L_x, L_y$ (lateral dimensions) and $L_z$ (thickness).
	
	The evaluation of the pinning parameters will then rely on a fitting procedure to provide the best fit to the experimentally observed spin wave spectra, where the particular$\ k$ values will be substituted in the resonance equation, such as expressed by Eq.~(\ref{eq:SWresInf})
	
	\begin{figure*}[ht]
		\center
		\includegraphics[width=13.5cm]{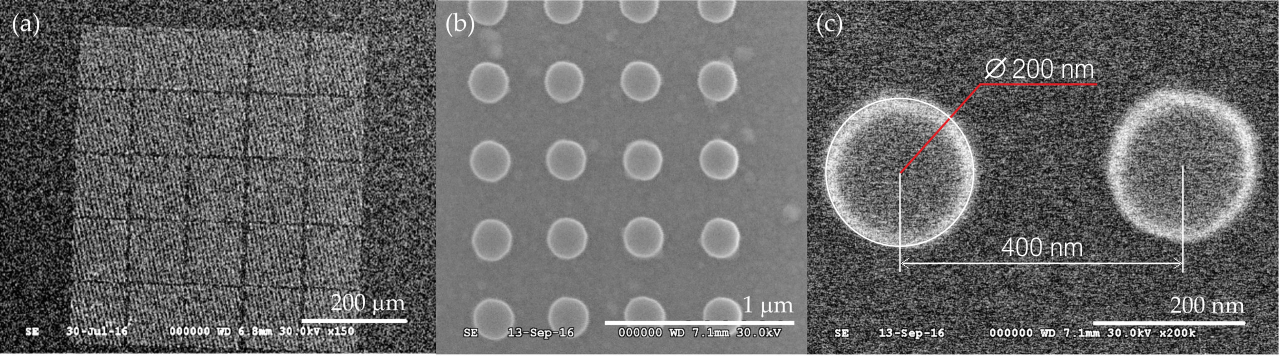} 
		\caption{\small Planar view SEM micrographs of a Co/Ag nanodot array at (a) 150$\times$, (b) 50,000$\times$ and (c) 200,000$\times$ magnification.}
		\label{Fig1}
	\end{figure*}
	
	\begin{figure}[]
		\centering
		\includegraphics[width=0.75\linewidth]{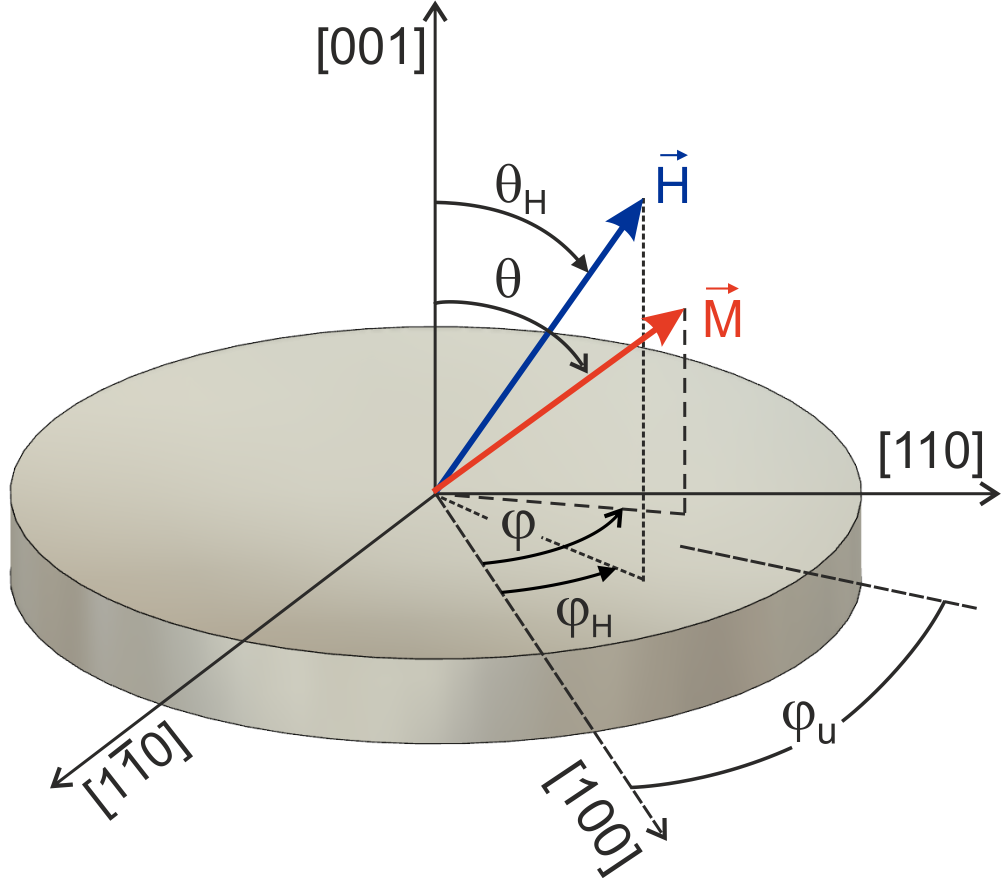}
		\caption{Definition of the azimuthal angles $\varphi$, $\varphi_H$, and $\varphi_u$ and the polar angles $\theta$ and $\theta_H$ with respect to the sample plane as well as the field and magnetization vectors (adapted from Ref.~\cite{Lenz2005}).}
		\label{geometryfmr}
	\end{figure}

	\section{Experimental Details}
	Samples of 50-nm-thick Co thin films were deposited on thermally oxidized silicon wafer substrates by ion-beam-assisted deposition (IBAD) \cite{Li2020}. During deposition, a Kaufmann ion source operating at \SI{800}{\volt} and \SI{7.5}{\mA} was used to sputter the Co target. The base vacuum and deposition pressure were 
	\SI{6.7d-5}{\pascal} and \SI{4d-3}{\pascal}, respectively. After the Co deposition, a 30-nm-thick Ag capping layer was subsequently deposited without breaking the vacuum. A combination of electron-beam lithography and ion milling was then used to pattern the Co/Ag thin film into a \qtyproduct{0.5 x 0.5}{\mm} nanodot array with an individual dot diameter of about \SI{200}{\nm} and a pitch of \SI{400}{\nm}. In Fig.~\ref{Fig1}, scanning electron microscopy (SEM) micrographs of a typical nanodot array at three different magnifications are shown.
	
	\section{Experimental Results and Analysis}
	\subsection{Vector Network Analyzer Ferromagnetic Resonance}
	
	Magnetization dynamics of the Co/Ag samples have been investigated by means of broadband vector network analyzer ferromagnetic resonance (VNA-FMR) at room temperature. For best signal-to-noise ratio, all measurements have been performed in field-sweep mode at fixed frequencies of up to \SI{40}{\GHz}. The samples were put face-side down onto an impedance-matched coplanar wave\-guide (CPW), which was connected by both nonmagnetic end-launch connectors and coaxial cables to a two-port VNA. A dc-bias magnetic field up to $\mu_0 H$ = \SI{2.2}{\tesla}, generated by an electromagnet, was applied perpendicular to the rf magnetic field $h_\textrm{rf}$ generated by the CPW. The microwave power was set to \SI{0}{dBm} for each measurement. A reference measurement with the sample saturated parallel to the rf magnetic field, such that any precession of the magnetization is completely suppressed, was subtracted from all acquired raw data sets. The magnitude of the background-subtracted forward transmission parameter $S_{21}$ was subsequently fitted to a complex Lorentzian to extract the corresponding resonance frequencies/fields as well as the peak-to-peak linewidth.  
	
	A total of four different measurement configurations have been investigated, in which either the polar angle $\theta_H$ or the azimuthal angle $\varphi_H$ of the dc bias magnetic field $H$ with respect to the sample and its anisotropy direction(s) was varied (see Fig.~\ref{geometryfmr}). The individual measurement configurations are the following: (i) in-plane $f(H)$ dependence: $H$ applied in the sample plane and parallel to the easy axis of the sample, (ii) out-of-plane $f(H)$ dependence: $H$ applied perpendicular to the sample plane, (iii) polar angular dependence: variation of $\theta_H$ between \ang{0} (out-of-plane) and \ang{90} (in-plane), and (iv) azimuthal angular dependence: full in-plane rotation of the sample in the field plane (\ang{0} $\leq \varphi_{H} \leq$ \ang{360}). By first performing the VNA-FMR measurements in all configurations in a specific order and then iteratively fitting all data sets, we are able to determine the magnetic parameters of the samples with best possible accuracy. The results are discussed in the remainder of this section.
	
	\begin{figure*}[ht]
		\center
		\includegraphics[width=0.675\linewidth]{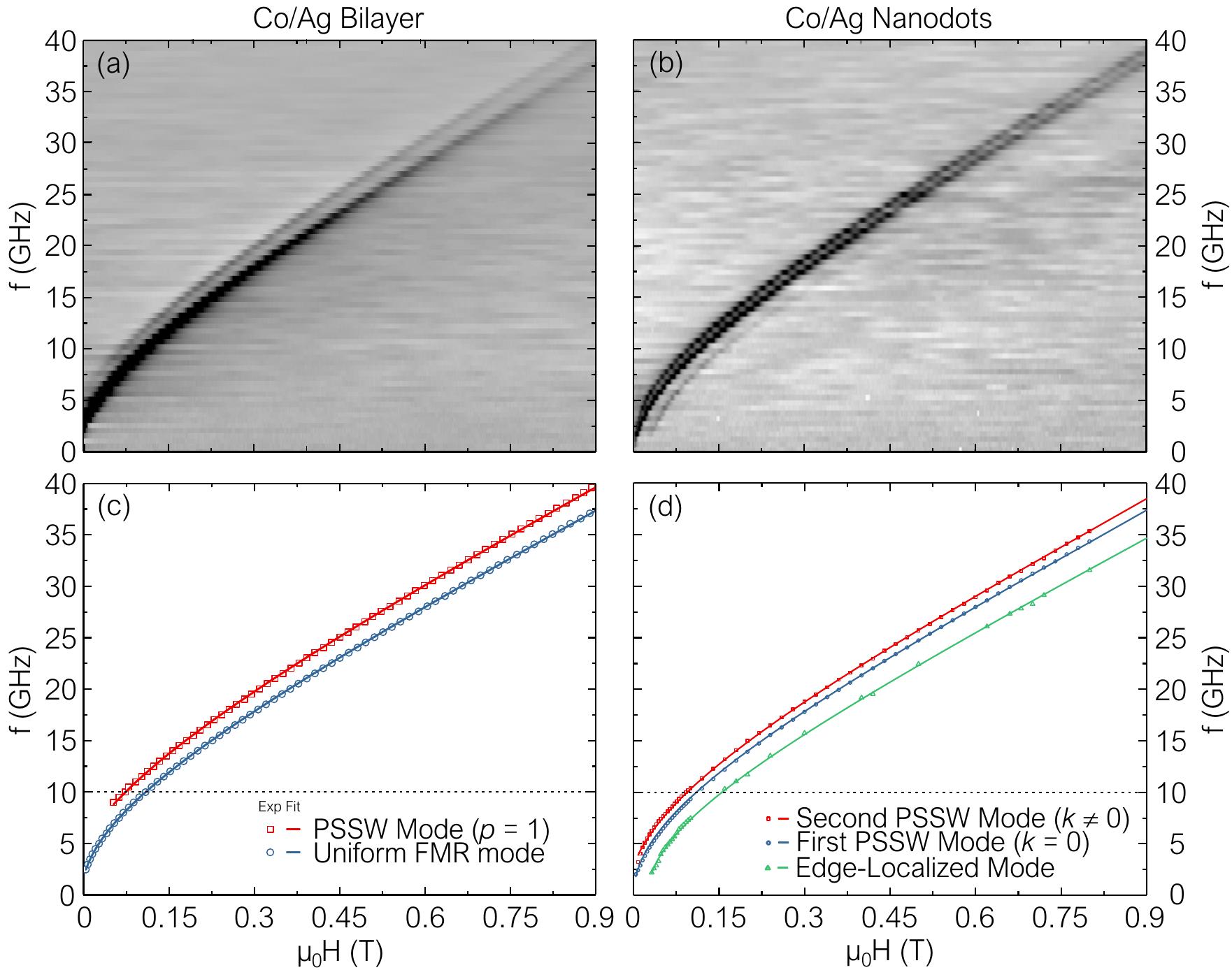} 
		\caption{\small In-plane easy-axis VNA-FMR spectra for the Co/Ag samples. Raw data for (a) the bilayer sample and (b) the nanostructured dots, extracted data (symbols) and corresponding fits (lines) for (c) the continuous film and (d) the nanodots.}
		\label{Fig3}
	\end{figure*}
	
	\subsection{In-plane VNA-FMR}
	\subsubsection{Continuous Film}
	In order to fully understand the effects of the nanostructuring on the FMR response of the nanodot array, we use the FMR spectra from the unpatterned reference film. This provides a reference of the dynamic response of the materials in question. We will proceed to analyze these spectra and then compare the results to the nanostructured samples. In Fig.~\ref{Fig3}, in-plane easy-axis VNA-FMR spectra for both the Co/Ag bilayer and nanostructured dots are shown \cite{Schmool2021}. 
	
	For the bilayer system, we observe two clear resonance lines, which we interpret as the uniform mode and the first PSSW mode with $p = 1$. The raw $f(H)$ characteristics and extracted data for the continuous film are illustrated in Figs.~\ref{Fig3}(a) and \ref{Fig3}(c), respectively. The uniform FMR mode is fitted to a modified version of the Kittel equation including the in-plane anisotropy field $H_K$:
	\begin{equation}
		\left(\frac{\omega }{\mu_0\gamma}\right)^2 = \left(H + H_K\right)\left(H + H_K + M_\textrm{eff}\right).
		\label{eq:Kittel+Aniso}
	\end{equation}
	Using a $g$-factor of 2.044, obtained from the out-of-plane FMR data (see Table \ref{tab:OOP}), we obtain $\mu_0 H_{K}$ = 4.5 mT and $\mu_0 M_\text{eff}$ = 0.97 T.
	For fitting the PSSW mode, we use a modified version of Eq.~(\ref{eq:SWresInf}), which for our purposes takes the form:
	
	\begin{equation}
		\left(\frac{\omega }{\mu_0\gamma}\right)^2 = (H + H_K + Dk^2)(H + H_K + Dk^2 + M_\textrm{s}).
		\label{eq:fmr1}
	\end{equation}
	We then apply Eq.~(\ref{eq:WV1Dip}) in the fitting process to determine the relevant wave vectors and hence the pinning parameter. Having used the following parameters for the fitting procedure: a $g$-factor of 2.044, an in-plane anisotropy field of $\mu_0 H_{K}$ = \SI{4.5}{\milli\tesla}, and an exchange stiffness constant $A_\textrm{ex}$ = \SI{1.0d-11}{\joule\per\meter}, which is reasonable for a 50-nm-thick Co film \cite{Eyrich2012}, we obtain $\mu_0 M_\textrm{s}$ = \SI{1.13}{\tesla}. For the pinning parameter, we obtain a value of $\delta = 0.50$ for the first PSSW mode with $p = 1$, where we have assumed a symmetric pinning with $\delta_1 = \delta_2 = \delta$. The fits shown in Fig.~\ref{Fig3}(c) are in excellent agreement with the experimental data and the physical parameters are consistent with the Co film studied. With regards to the pinning conditions, we note that the value for $\delta$ indicates intermediate pinning on the upper and lower film interfaces.\\\\
	
	\subsubsection{Nanodot Array}
	\label{Nanodot_array}
	
	The raw and extracted in-plane VNA-FMR data of the nanodot array is shown in Figs.~\ref{Fig3}(b) and \ref{Fig3}(d), respectively. At first glance, we note that there are two dominant modes with roughly the same intensity and which appear to be modified with respect to the resonances observed in the continuous film. They appear closer together than in the single Co layer. In addition to these modes, we note a further resonance line of weaker intensity on the high-field (low-frequency) side of the main resonances. In the extracted data, Fig.~\ref{Fig3}(d), we see that the stronger resonance lines are roughly parallel at higher magnetic field values and appear to merge at low fields. The lower resonance line, shown in green, has a slightly lower gradient in the linear portion of the $f(H)$ characteristics. 
	
	Since the main resonance lines appear to be related to those of the continuous thin film, we can assume that the upper and lower boundary conditions should be the same. This will mean that we can concentrate on the effects of the patterning, which restrict the lateral dimensions of the nanostructures, in interpreting the experimental data. The first line that we can consider is the blue line, which appears to be very close to the corresponding blue line (uniform mode) in the continuous film data. The red line is somewhat shifted to smaller resonance frequencies with respect to the continuous film (PSSW mode), but otherwise appears to be the same mode, which can be accounted for by a modified wave vector for this mode. Fitting these two lines can be based on the same principles as that for the thin film sample. In the latter, we considered that for a thin film the in-plane wave vectors should be zero since the film is effectively infinite in the film plane: $k_\textrm{x,y} = (p_\textrm{x,y} - \delta_\textrm{x,y} ) \pi^2 /L_\textrm{x,y} ^2 = 0$ since $L_\textrm{x,y} \rightarrow \infty$ and thus $k_{||} = 0$, see Eq.~(\ref{eq:WV3Dasym}). For the case of our circular nanodots, the lateral dimensions, corresponding to the \SI{200}{\nm} diameter, can give rise to pinning conditions at the dot edges, and thus, we should consider the three-dimensional aspect of magnetic confinement. Using Eq.~(\ref{eq:WV3Dasym}) for symmetric pinning, we consider the dot geometry and the edge pinning conditions, which we can express in a similar manner to the thin film case, from which we can write:
	\begin{widetext}
		\begin{equation}
			k^2_{pqr} = k_{\zeta}^2 + \left(\frac{p' \pi}{L}\right)^2 = \left[ (p - \delta_\textrm{d})^2 + (q - \delta_\textrm{d})^2\right] \left(\frac{ \pi}{d}\right)^2 +(r - \delta_\textrm{L})^2\left(\frac{\pi}{L}\right)^2, 
			\label{eq:WVgen}
		\end{equation}
	\end{widetext}
	where $k_{\zeta}^2 = k_\textrm{y}^2 + k_\textrm{z}^2 $ is the in-plane component of the wave vector, $\delta_\textrm{d}$ denotes the edge pinning conditions and $\delta_\textrm{L}$ those of the upper and lower interfaces, $d$ is the dot diameter, and $L$ the film thickness. In considering the three-dimensional case, we need to account for the mode numbers in three directions, as represented by the integers $p, q$, and $r$. We note that Eq.~(\ref{eq:WVgen}) considers that the pinning is equivalent in both lateral directions and is symmetric. This is justified for the current geometry, since there is no reason to assume that there should be any variation of the pinning at the edges of the dot structure. In our calculations, we consider that the in-plane anisotropy is weak and that inter-dot (dipole-dipole) interactions are sufficiently weak so that we can neglect their effect. 
	
	By taking the specific conditions of the nanodots into account, we note that the thickness of the dots (as that of the continuous layer) is $L$ = \SI{50}{\nm}, while the dots have lateral dimensions (diameter) of $d$ = \SI{200}{\nm}. As a first consideration, we can compare the factors in the wave vectors, $n_1 = \pi^2/L^2$ and $n_2 = \pi^2/d^2$. These yield $n_1 \simeq$ \SI{3.95d-15}{m^{-2}} and $n_2 \simeq$ \SI{2.47d-14}{m^{-2}}. This indicates that the thickness dimension will still dominate the wave vector, but modifications should be accounted for due to the lateral dimensions. 
	\newpage
	Comparing the dispersion relations for the continuous thin film and the nanostructured sample, see the red and blue lines in Figs.~\ref{Fig3}(c) and \ref{Fig3}(d), we see that the two resonances lines from the former are modified in the latter. They are closer together in the nanodot array, and on closer inspection, the blue line is virtually unchanged in the two samples. This means that the second mode is different in nature and must be attributed to the spin-wave mode wave vector. In the case of the thin film, we interpreted this as arising from the next excited spin wave, and here we must similarly attribute the second mode (red) to the next excited spin wave. From the above analysis, we can do this by \mbox{assigning} the relevant mode numbers for the two spin-wave resonances. Given that the thin film analysis indicates that the boundary spins are close to a free pinning (or unpinned) condition, we can look for solutions based on the resonance equations (\ref{eq:DispGen}) and (\ref{eq:Fnn}) given in the theory section. The latter equation should be expressed in the form:
	\begin{widetext}
		\begin{equation}	
			P_{pq} = \frac{k_r^2}{k_{q}^2} \delta_{pq} - \frac{k_{r}^4}{k_{p}^2 k_{q}^2}F_p \frac{1}{\sqrt{(1+\delta_{0p})(1+\delta_{0q})}}\left( \frac{1 + (-1)^{p+q} }{2}\right)
			\label{eq:Pnn}
		\end{equation}
	\end{widetext}
	for this case, see Ref.~\cite{Kalinikos1986}, where
	\begin{equation}	
		F_p = \frac{2}{k_r L} \bigl[1 - (-1)^p e^{-k_r L}\bigr],
		\label{eq:Pnn'}
	\end{equation}
	$\delta_{pq}$ is the Kronecker delta and $k_{p}^2 = k_r^2 + \kappa_p^2 = k_r^2 + (p\pi/L)^2$. For the lowest lying mode, with $p, q = 0$, we obtain $P_{00} \simeq 1$ and$\ F_{00} \simeq \sin^2 \theta $ such that resonance equation (\ref{eq:Fnn}) yields:
			\begin{equation}	
			\omega_{pqr}^2\!=\!(\omega_H\!+\!\omega_D k_{pqr}^2)(\omega_H\!+\!\omega_D k_{pqr}^2\!+\!\omega_M \sin^2 \theta_k),
			\label{eq:resllm}
		\end{equation}
	which is essentially the uniform mode as expressed by the Kittel-Herring equation (\ref{eq:SWresInf}). The next excited mode for$\ p, q = 0,1$ (or 1, 0), yields $P_{01} = P_{10}  \simeq 0$ and$\ F_{01}  \simeq1$ and so on. It is worth noting that we consider $r = 0$ here, since this will correspond to the thin-film-like geometry, where $k_r$ is the perpendicular component of the wave vector. We note that the parameters $\omega_H, \omega_M$, and $\omega_D$ are given earlier. 
	
\begin{figure}
		\centering
		\includegraphics[width=0.98\linewidth]{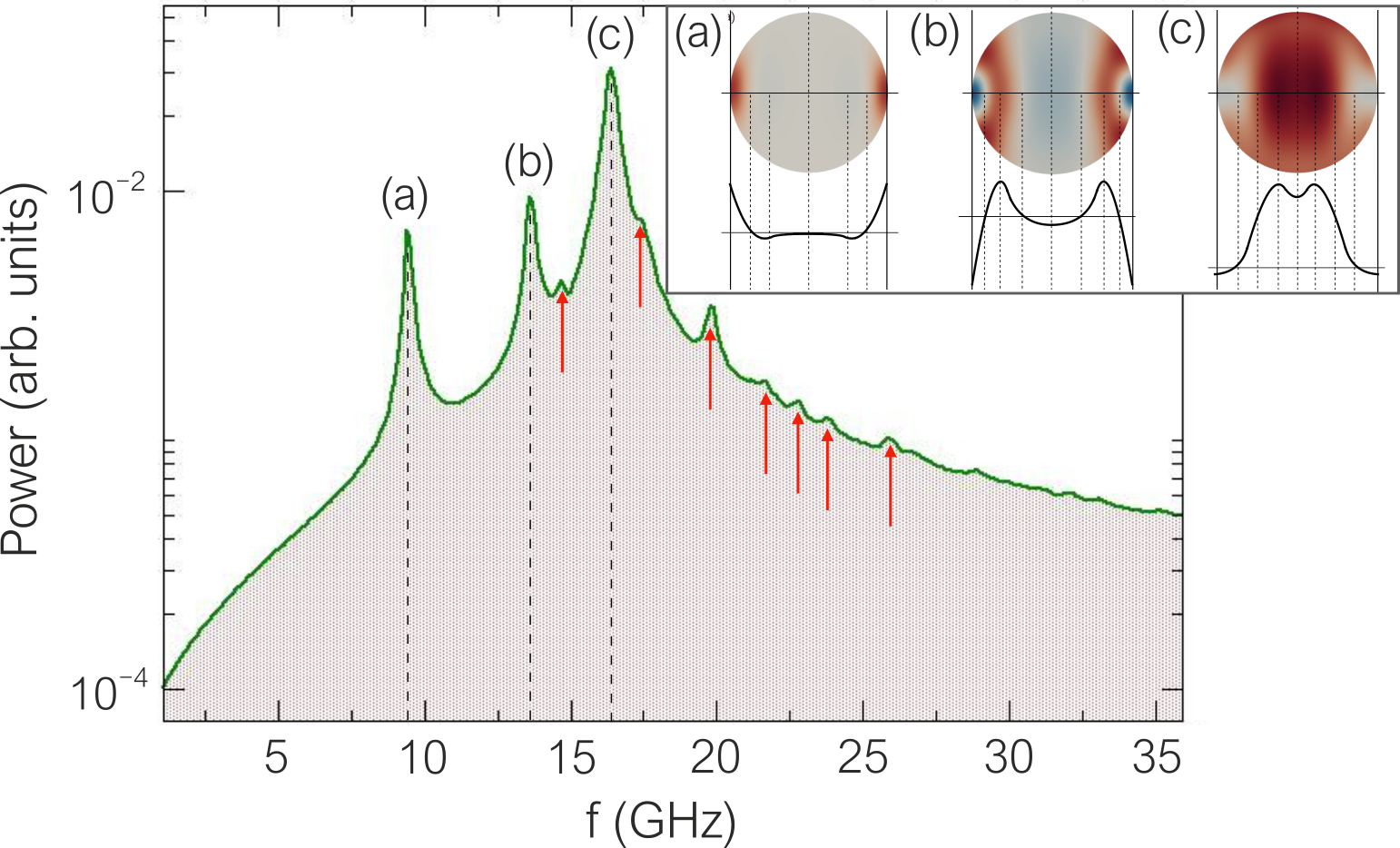}
		\caption{Simulated absorption-power spectrum for the Co nanodot structure as a function of the excitation frequency for an in-plane applied magnetic field $\mu_0H$ = \SI{0.3}{\tesla}. The modal patterns are given for the three most intense absorption lines, labeled as (a), (b), and (c). Below each modal pattern, a schematic representation of the mode profile (cross-section) is shown.}
		\label{SimSpectr}
	\end{figure}

	We now have the basic necessary elements to evaluate the observed spectra. However, the complexity of the analysis means that we require some further input into identifying the possible excitation modes responsible for the observed resonance peaks. 
	To assist this process, we have used micromagnetic simulations of the nanodot structure. For a full discussion of the simulations see Appendix A. 
	These simulations provide us with the expected power spectrum for the resonance modes of the nanostructure with the corresponding size, shape and material parameters, under specific conditions of the applied static magnetic field and excitation frequency. As an example, we illustrate in Fig.~\ref{SimSpectr}, the simulated power spectrum for the basic nanodot structure of this study. We note that the power scale is logarithmic, meaning that the main peak at a frequency of 16.3 GHz is significantly stronger than the other modes.
	Considering the three principal modes, at frequencies of \SI{9.4}{\GHz}, \SI{13.6}{\GHz}, and \SI{16.3}{\GHz}, we can map the corresponding modal patterns for these excitations, which are illustrated in the inset of Fig.~\ref{SimSpectr}. 
	
	We now compare between the simulated and observed spectra to try to allocate the most likely mode patterns. We must, however, bear in mind that we cannot realistically expect an exact one-to-one correspondence and we use this procedure with caution, where we note that small modifications in the model structure and parameters can give rise to strong modifications in the simulated spectra and excitation modes. First of all, we note that the simulated modes, for an applied field $\mu_0H$ = \SI{0.3}{\tesla}, at \SI{9.4}{\GHz} and \SI{13.6}{\GHz} have modal patterns indicating an edge-localized character. In this case, both modes have excitations, which are concentrated at the edge of the nanodot, while the center of the dots shows no spin precession. We note that these modes, due to the localized nature of the excitation, will have frequencies below the theoretical uniform mode and excitations fields above the uniform resonance field \cite{Puszkarski1979}. Indeed, the more localized the mode, the larger the frequency shift with respect to the uniform mode. This is seen to be the case, where the mode with \SI{9.4}{\GHz} is further away from the uniform mode than the resonance with \SI{13.6}{\GHz}. We note that the former only has spin-precessional activity at the extremes (left and right) of the nanodot, see Fig.~\ref{SimSpectr}(a), while in the latter, the precessional motion extends further towards, but not at the center of the dot, Fig.~\ref{SimSpectr}(b). This means that the latter mode is less localized than the former. The form of the localized mode (a) is very similar to the pattern reported in Ref.~\cite{Zivieri2006}. 
	
\begin{figure}[t!]
		\centering
		\includegraphics[width=0.98\linewidth]{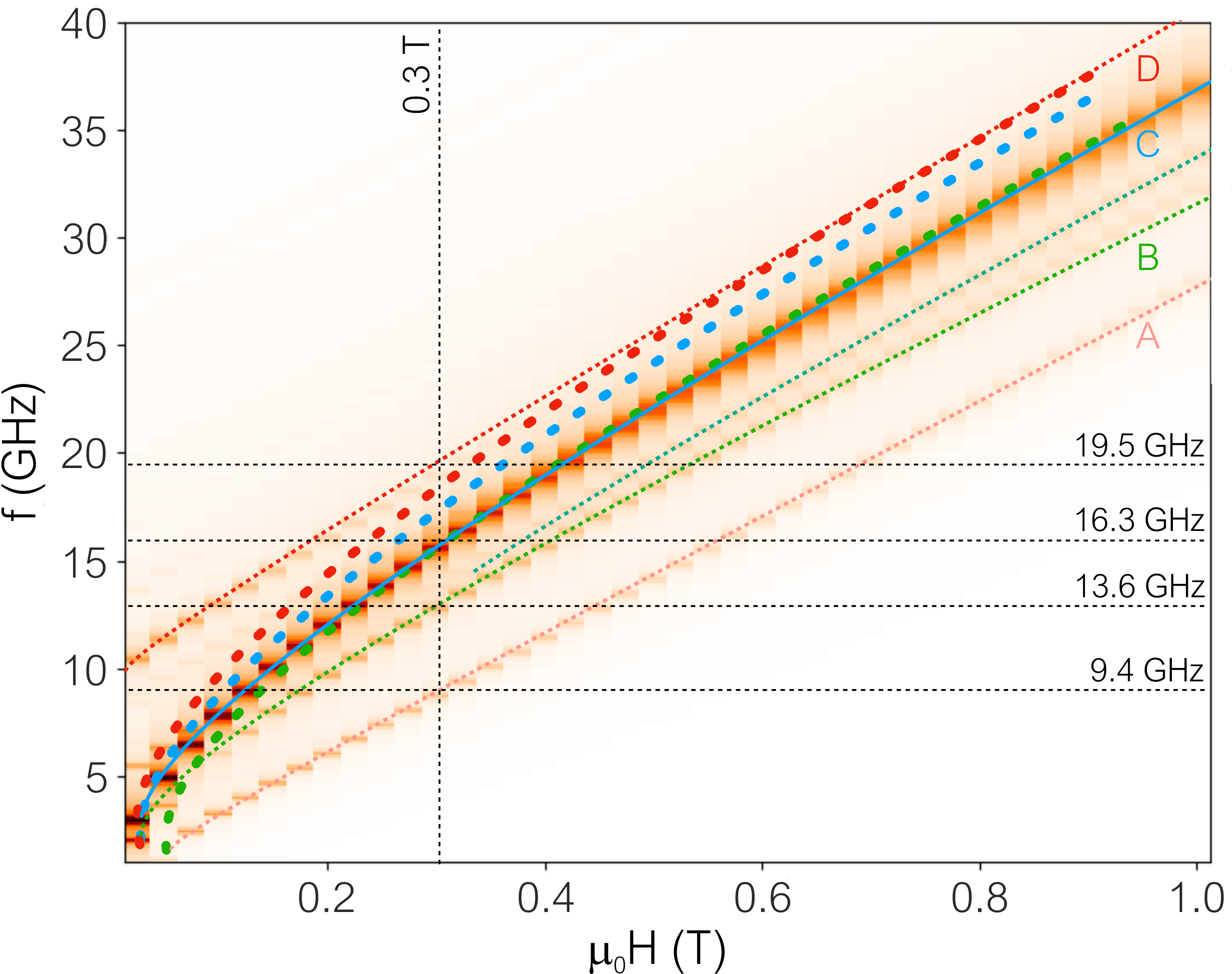}
		\caption{$f(H)$ plot of the simulated in-plane spectra. Principal lines are indicated as lines A through D. See Fig.~\ref{fig:Simulations} in Appendix A for the original data without added lines. The spectrum illustrated in Fig.~\ref{SimSpectr}, for an applied field $\mu_0H$ = \SI{0.3}{\tesla}, is indicated with the vertical line at this field value and the frequencies of the principal modes are also shown explicitly. We further show, for comparison, the lines (dotted) corresponding to the experimental variation for the three resonances found with colors corresponding to those in the experimental curves of Fig.~\ref{Fig3}.}
		\label{fHvarn}
	\end{figure}

	The most intense mode in the simulated spectrum is observed at \SI{16.3}{\GHz} for $\mu_0H$ = \SI{0.3}{\tesla}. The shape of this mode, as shown in Fig.~\ref{SimSpectr}(c), is the fundamental mode for the specific boundary condition of the nanodots. We note that this fundamental mode is not strictly speaking the uniform mode, which consists of the pure FMR mode, where all spins in the magnetic object precess in unison. Hence, it is sometimes termed "center mode". This applies to both bulk spins as well as surface and boundary spins and can be considered as a special case. In Fig.~\ref{fHvarn}, we illustrate the full simulated $f(H)$ variation in the corresponding range for the experimental measurements. The lines are named A through D. Note that lines A, B and C correspond to the modes indicated as (a), (b) and (c) in Fig.~\ref{SimSpectr}. We have added lines as a guide to the eye since some of the modes are of weak intensity. The principal resonance line here is indicated as line C. We have further added the line corresponding to the experimental data for the principal resonance line (red dotted line), taken from Fig.~\ref{Fig3}. There is an overall good qualitative agreement between the simulated and experimental data for this line.
	
	We will now consider the nature of the non-principal modes and attempt a comparison with the experimental spectra and thus try to interpret the experimental data. In the simulated spectra, lines A and B are seen to occur at frequencies below the principal resonance, indicative of localized standing spin wave modes, as is indeed illustrated from the modal patterns (a) and (b) in Fig.~\ref{SimSpectr}. The degree of localization is related to the mode profile and the extent of the spin precession towards the interior of the magnetic body. Greater localization is associated with a larger shift of the resonance frequency from that of the uniform resonance mode (with a zero wave vector, $k= 0$). It is worth noting that such localized modes have an imaginary wave vector of the form $k = i\tau$. For a more in-depth discussion of the positioning of the resonance lines, see Appendix B.
	In this case, we see that mode A is more localized than mode B, and as such is further away from the principal resonance line (even if this is not a uniform resonance). In relation to the experimental measurement, the weakest line (in green), see Figs.~\ref{Fig3}(b) and \ref{Fig3}(d), is clearly of this nature and appears as a weak resonance above the principal resonance. This line will then follow a resonance equation of the form (see Appendix B):
	\begin{equation}
		\left(\frac{\omega_{\tau}}{\mu_0\gamma}\right)^2 = \left(\frac{\omega_\text{FMR} }{\mu_0\gamma}\right)^2 -D\tau^2 (2H +  M_\textrm{s} ) + (D\tau^2)^2.
		\label{eq:LocResEq}
	\end{equation}
	
	The fit for the green line in Fig.~\ref{Fig3}(d) has been performed for a value of $\tau$ = \SI{2.85d7}{m^{-1}}, or $k^2$ = \SI{-8.1225d14}{m^{-2}}. Since this mode does not lie far from the principal resonance, the localization is not very strong and the mode profile is expected to be of a form close to that illustrated for mode (b) in Fig.~\ref{SimSpectr}.  That said, we cannot exclude \textit{a priori} other localized modes. This is because mode intensities are proportional to the transversal magnetization, meaning that strongly localized modes, which lie further away from the principal (and fundamental) modes, will become rather weak in intensity and can be difficult to observe.
	
	Let us now consider the two bulk (VSSW) resonances. These correspond to the blue and red lines in Fig.~\ref{Fig3}(d). While the simulations do not give a good quantitative agreement for these lines, we can still consider a modal pattern, which resembles the simulated VSSW modes for the nanodot. An approximation of this mode structure may be considered in which the boundary condition along the dot is accounted for by a perfect pinning, while that across the dot has free boundaries, or perfect freedom. According to Eq.~\ref{eq:WVgen}, we can look for modes with wave vectors of the form:
	\begin{equation}	
		k^2_{pqr} \simeq \bigl[p ^2 + (q - 1)^2 \bigr] \left(\frac{ \pi}{d}\right)^2 +(r -1)^2\left(\frac{\pi}{L}\right)^2. 
		\label{eq:WVmixedbound}
	\end{equation}
	
	Based on this expression, we can calculate the lowest order mode numbers. Since we will consider only the fundamental excitation in the direction perpendicular to the plane of the dots (which is a valid assumption since these will give the lowest lying VSSW modes for the structure), we will set $r=1$ since we are considering only the lowest lying modes. A calculation of the lowest lying modes, based on Eq.~(\ref{eq:WVmixedbound}), is given in Appendix C. Our spin-wave spectrum will reflect the mode order in terms of their relative energies and subsequent frequencies as well as the degeneracy of the modes, see Table \ref{tab:Mode numbers}.
	
	Let us now consider the resonance equation for the volume modes. Using expressions (\ref{eq:Pnn}) and (\ref{eq:Pnn'}), and taking into account that for the lowest lying modes we have $r=1$, we find $P_{pq}=0$ and from expression (\ref{eq:Fnn}) $F_{nn}(k_\zeta L) = \sin^2\theta$. In this case, the resonance equation takes the same form as illustrated in Eq.~(\ref{eq:resllm}). We can now rewrite this equation in the form:
	\begin{widetext}
		\begin{equation}
			\left(\frac{\omega_{pqr}}{\mu_0\gamma}\right)^2 = \left(\frac{\omega_\text{FMR} }{\mu_0\gamma}\right)^2 +Dk^2_{pqr} (2 H + M_\textrm{s} \sin^2\theta_k) + (Dk^2_{pqr})^2.
			\label{eq:VSSWEq}
		\end{equation}
	\end{widetext}
	This is again a convenient form, since it allows us to compare the line shifts with respect to the uniform mode. In this case, we see that the modes will be frequency up-shifted with respect to this mode. From the table in Appendix C, we see that the lowest lying modes have $k^2_{011} = 0$, for the 011 mode, $k^2_{001} = k^2_{021} = k^2_{111} = n_2$ and $k^2_{101} = k^2_{121} = 2n_2$. This means that for this particular set of values, the lowest resonance mode looks like a uniform mode, and once again is the consequence of the choice of pinning conditions. Above this we have a triplet state, followed by a doublet. 
	
	\begin{table*}[t]
		\centering
		\caption{Calculated frequency spectrum for the Co nanodots based on Eq.~(\ref{eq:Freqpqr}) at an applied in-plane field $\mu_0 H$ = \SI{0.3}{\tesla}.}
		\label{tab:GenFreqSpectr}
		\begin{ruledtabular}
			\begin{tabular}{cccc}
				Modes ($pqr$) & {$k^2$ ($\times 10^{14}$ m$^{-2}$)} & {Calculated frequency (GHz)} & {Experimental frequency (GHz)} \\
				\colrule
				011 & 0 & 17.48\;{(uniform mode)} & 17.52\\
				001, 111, 021 & 2.47 & 17.70 &  \\		
				101, 121 & 4.94 & 17.92 &  \\	
				211, 031 & 9.88 & 18.36 &  \\	
				201, 221, 131 & 12.35 & 18.57 & 18.65 \\
				311, 041 & 22.23 & 19.43 &  \\
				301, ... & 24.70 & 19.64 &  \\
			\end{tabular}
		\end{ruledtabular}
	\end{table*}
	
	\pagebreak
	For the fundamental mode, we can express the resonance equation in a simplified form as:
	\begin{equation}
		\omega_\text{FMR} =\gamma \sqrt{\mu_0H ( \mu_0H + \mu_0M_\textrm{s})}
		\label{eq:fmrfund}
	\end{equation}
	
	At an applied field of $\mu_0 H$ = \SI{0.3}{\tesla}, we can calculate the expected uniform-like ferromagnetic resonance frequency as $f_\text{FMR} =\omega_\text{FMR} /2 \pi$ = \SI{17.48}{\GHz}, where we have used $g= 2.04$, $\gamma$ = \SI{179.4}{\GHz.\tesla^{-1}} and $\mu_0M_\textrm{s}$ = \SI{1.08}{\tesla}, as obtained from the thin film sample. Comparing this value to the one observed experimentally, see Fig.~\ref{Fig3} where $f_\text{exp}$ = \SI{17.52}{\GHz}, we have an excellent agreement. 
	
	From the expected SW spectrum, see Appendix C, we can now evaluate the frequencies of the lowest lying modes, which are calculated from Eq.~(\ref{eq:VSSWEq}) and which we can express in a modified form as:
	\begin{widetext}
		\begin{equation}
			f_{pqr}^2 =  f_\text{FMR}^2 + \frac{(\mu_0\gamma)^2}{4\pi^2} \bigl[D k_{pqr}^2 (2 H +  M_\textrm{s} ) + (D k_{pqr}^2)^2 \bigr],
			\label{eq:Freqpqr}
		\end{equation}
	\end{widetext}
	where we take $\theta_k = \ang{90}$ for the in-plane geometry and configuration used in the experiment.
	Using the same magnetic constants as expressed above, and for an applied field $\mu_0 H$ = \SI{0.3}{\tesla}, we generate the frequency spectrum given in Table \ref{tab:GenFreqSpectr}.
	
	We also note in Table \ref{tab:GenFreqSpectr} the values of the experimental frequencies, where the best fit to the calculated values corresponds to the 201, 131 modes. While we are not claiming an exact agreement between experiment and theory, we can see that due to the closely spaced frequency values generated from the calculations, we have a plausible agreement, though not all modes are observed experimentally. It is of course pertinent to ask why some modes are not observed. Some modes will not be excited, since the modes will have a zero component of the transverse field, which excludes them from the measured excitation spectrum. Given that some of the predicted modes are only separated by about \SI{0.2}{\GHz}, this may also contribute to the fact that some modes cannot be distinguished from each other in the experiment. We further note in passing that additional excitation modes may also be present in the simulated spectra, though the weakness of the intensities means that we are not entirely confident of the reliability of such an interpretation. Small discrepancies between calculated and experimental values can be expected due to the simplifying assumptions that were made in approximating the $\delta$ values for the pinning conditions used in the wave vector, Eq.~(\ref{eq:WVmixedbound}).
	\begin{figure*}
		\centering
		\includegraphics[width=0.7\textwidth]{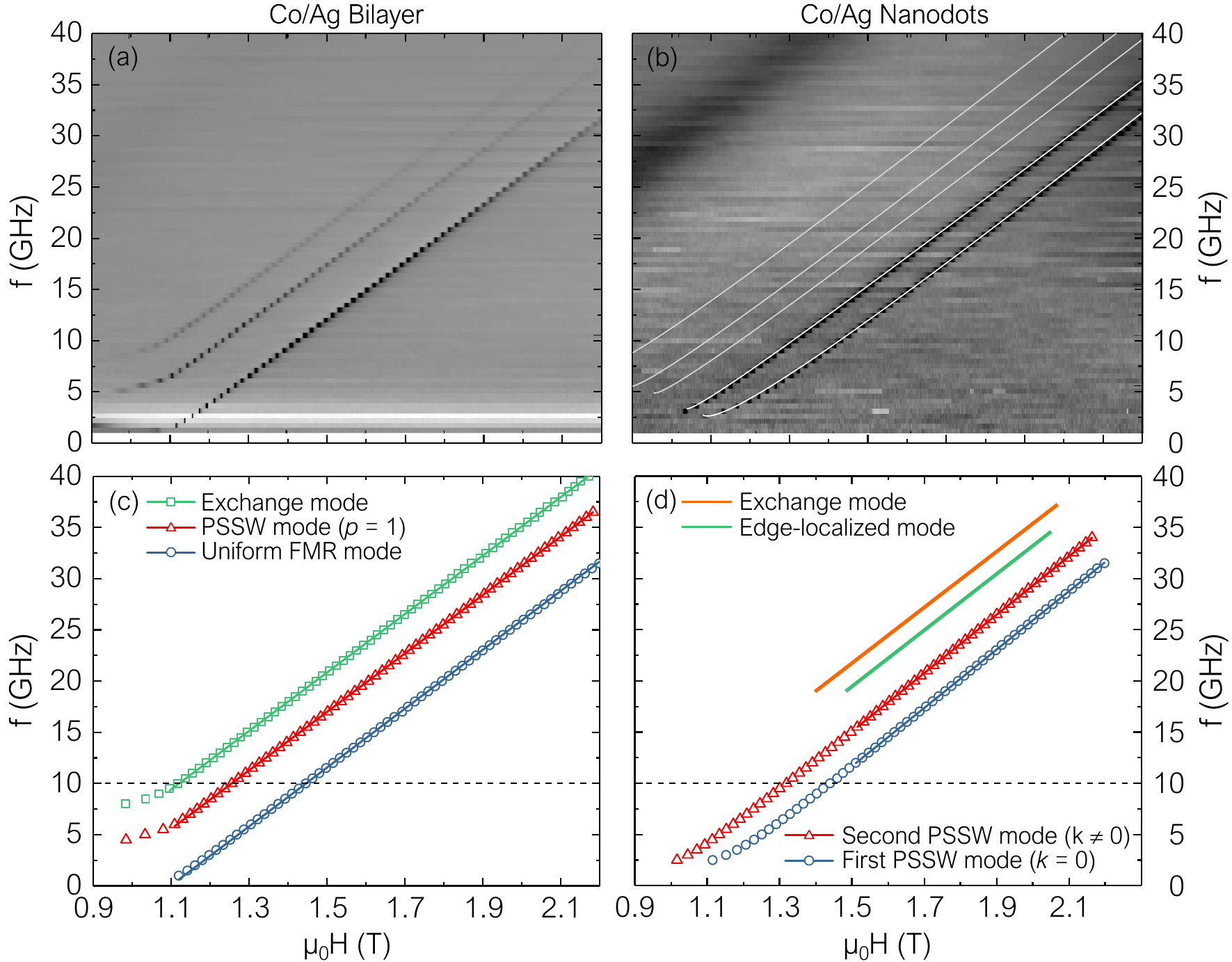}
		\caption{Out-out-plane $f(H)$ dependence for the Co/Ag bilayer (a) and the nanodots (b). The extracted resonance modes and corresponding fits to Eq.~(\ref{eq:O-FMR}) are depicted in (c) and (d), respectively.}
		\label{fig:OOP-FMR}
	\end{figure*}
	\vspace{1cm}
	\subsection{Out-of-plane VNA-FMR}
	
	In the out-of-plane (OOP) geometry, the $f(H)$ dependence shown in Fig.~\ref{fig:OOP-FMR} was measured in field-sweep mode as $f$ was gradually increased from \qtyrange{0.5}{40}{\GHz} in steps of \SI{0.5}{\GHz}, while the dc magnetic field $H$, applied perpendicular to the sample plane, was swept from \qtyrange{0}{2.2}{\tesla}. 
	
	The spectrum of the bilayer sample [Fig.~\ref{fig:OOP-FMR}(a)] contains three clearly visible resonance lines, whereas only two modes are clearly visible in the nanodot array spectrum [Fig.~\ref{fig:OOP-FMR}(b)]. However, the latter contains at least two, if not even three, additional modes of very weak intensity, which cannot be extracted from the measured data points, but can only be redrawn in the corresponding absorption plot as depicted in Fig.~\ref{fig:OOP-FMR}(b). In the bilayer sample, the mode intensity of the individual resonances lines decreases with both increasing frequency and field. In the nanodot array, however, each mode maintains its intensity throughout the entire field and frequency range. The frequency spacing between the individual modes exhibits a mixed behavior, that originates most likely from different boundary conditions for each set of modes.
	
	From the corresponding resonance equation:
	\begin{widetext}
		\begin{equation}
			\bigg(\frac{\omega}{\gamma}\bigg)^2 = \bigg(\mu_0 H-\mu_0 M_\textrm{eff}+\frac{2K_{4\perp}}{M_s}\bigg) \cdot \bigg(\mu_0 H-\mu_0 M_\textrm{eff}+\frac{2K_{4\perp}}{M_s}-\frac{2K_{2\parallel}}{M_s}\bigg)
			\label{eq:O-FMR}
		\end{equation}
	\end{widetext}
	it can be seen that in this measurement geometry the $f(H)$ dependence is practically linear in the high-field regime and that the effective magnetization $M_\textrm{eff}$ can be determined from the intercept with the field axis. The deviation from this linear dependence of the resonance frequency at fields smaller than \SI{1.2}{\tesla} and \SI{1.3}{\tesla} for the bilayer and the nanodots, respectively, stems from the fact that at those fields the magnetization is not yet fully perpendicular to the sample plane. The magnetic parameters obtained by fitting the high intensity modes in Figs.~\ref{fig:OOP-FMR}(c) and \ref{fig:OOP-FMR}(d) to Eq.~(\ref{eq:O-FMR}) are given in Table \ref{tab:OOP}.
	
\begin{figure}[ht]
		\centering
		\includegraphics[width=0.92\linewidth]{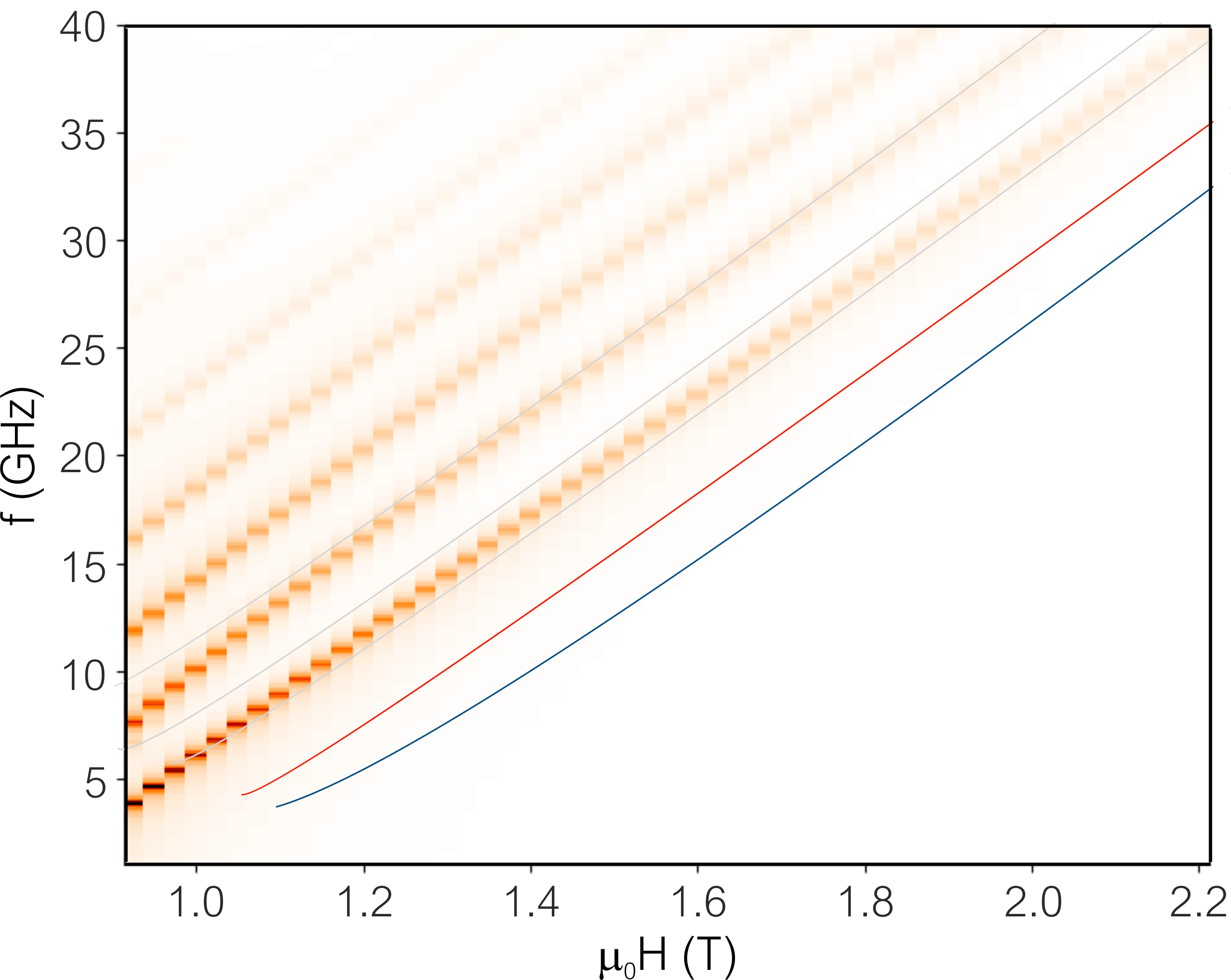}
		\caption{Simulated $f(H)$ spectrum for the out-of-plane \mbox{geometry}. The lines correspond to the measured modes in Fig.~\ref{fig:OOP-FMR}b.}
		\label{fig:oopsimfmr}
	\end{figure}

	In Fig.~\ref{fig:oopsimfmr}, the simulated $f(H)$ spectrum for the OOP geometry together with an overlay of the five modes having the lowest resonance frequencies from the corresponding experimental spectrum [Fig.~\ref{fig:OOP-FMR}(b)], indicated by the white, red, and blue lines, respectively, are depicted. Compared to the measured FMR spectrum, the simulated spectrum shows an even higher number of modes---at least six---and the corresponding mode intensities quickly decreases with increasing resonance field. It is obvious that there is a rather large mismatch between the calculated and measured frequencies, which presumably originates from the fact that the simulation was performed using only a single nanodot without considering the dipolar coupling between adjacent nanodots in the nanostructured array. 
	
	\begin{table}[b]
		\centering
		\caption{Magnetic parameters obtained by fitting the out-of-plane VNA-FMR data.}
		\label{tab:OOP}
		\begin{ruledtabular}
			\begin{tabular}{lcc}
				\multicolumn{1}{c}{} & \multicolumn{1}{c}{Bilayer}& \multicolumn{1}{c}{Nanodots} \\
				\cmidrule{2-2} \cmidrule{3-3} 
				Parameter & Uniform mode &  PSSW mode 1 \\		 
				\cmidrule{1-1} \cmidrule{2-2} \cmidrule{3-3}
				$g$-factor & 2.044 & 2.036 \\
				$\mu_0 M_\textrm{eff}$ (T) & 1.096 & 1.098 \\		
			\end{tabular}
		\end{ruledtabular}
	\end{table}
	
	\begin{figure}[h]
		\centering
		\includegraphics[width=0.925\linewidth]{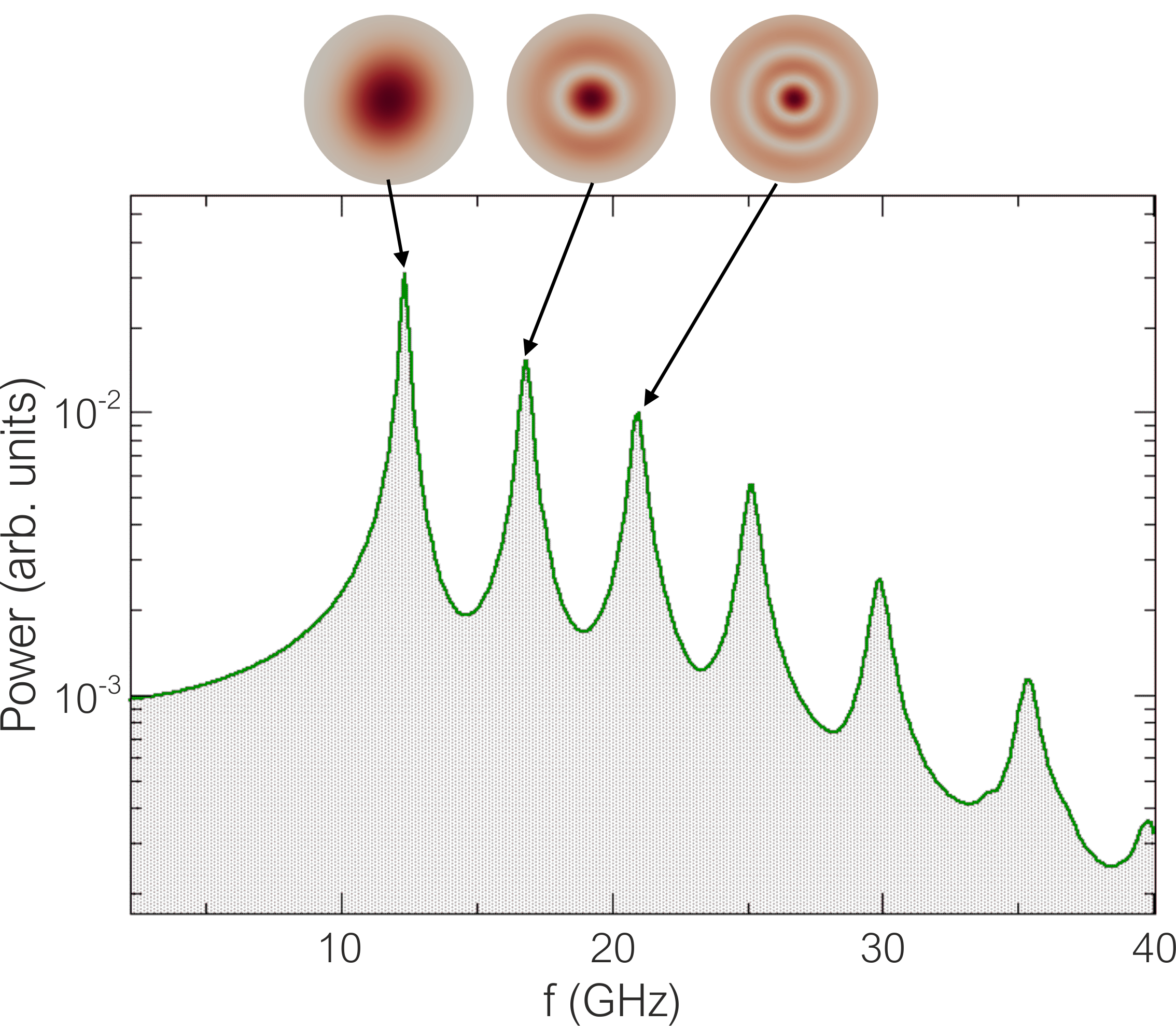}
		\caption{Absorption power spectrum of a single Co nanodot as a function of the excitation frequency $f$ at an applied out-of-plane field $\mu_0H$ = \SI{1.2}{\tesla}. Radial profiles of the first three standing spin-wave modes described by zeroth-order Bessel functions are shown in small insets.}
		\label{fig:newsimoop1}
	\end{figure}
	
	In Fig.~\ref{fig:newsimoop1}, the simulated power spectrum density of a single nanodot in the out-of-plane geometry as a function of the excitation frequency $f$ at $\mu_0H$ = \SI{1.2}{\tesla} is depicted. It can be seen that the amplitudes of the modes quickly decrease with increasing resonance frequency, i.e., with increasing mode number, as indicated by the logarithmic power scale, and that the frequency spacing between two adjacent modes increases with increasing mode number. For the three modes having the lowest resonance frequencies, small insets show the corresponding radial spin wave patterns. We note that the modal patterns shown are in excellent agreement with those determined by Dobrovolskiy et \textit{al.}~\cite{Dobrovolskiy2020}. \\
	
	\subsection{Azimuthal Angle-Dependent VNA-FMR}
	
\begin{figure}[h!]
		\centering
		\includegraphics[width=0.457\textwidth]{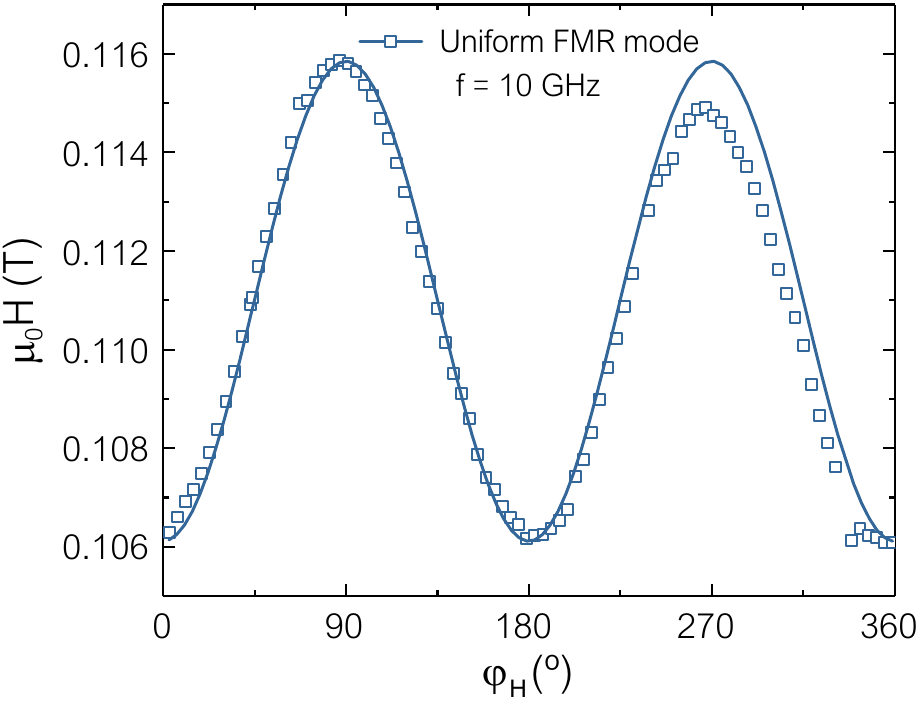}
		\caption{Azimuthal angular dependence of the FMR of the Co/Ag bilayer measured at $f$ = \SI{10}{\GHz}.}
		\label{fig:AZI-FMR}
	\end{figure}

	The azimuthal angular dependence of the FMR was measured at a fixed frequency of $f$ = \SI{10}{\GHz} in steps of \ang{5} for \ang{0} $\leq \varphi_H \leq$ \ang{360}. The corresponding $H(\varphi_H)$ dependence of the Co/Ag bilayer is depicted in Fig.~\ref{fig:AZI-FMR}. 
	
	The spectrum of the bilayer sample contains a single resonance, the uniform mode, exhibiting a two-fold symmetry with both two distinct maxima and minima across the entire range of $\varphi_H$, indicative of a uniaxial anisotropy. However, from the fact that the maxima at $\varphi_H$ = \ang{90} and $\varphi_H$ = \ang{270} have different values, we conclude that there is also a small unidirectional contribution of about \SI{1}{mT} to the in-plane magnetic anisotropy like in exchange-biased systems. However, we did not observe any frequency shift in the in-plane FMR spectra as these were measured only with the dc-bias field applied the along the easy axis, whereas the exchange bias field is oriented along the hard axis.  
	
	\begin{figure*}[ht]
		\centering
		\includegraphics[width=0.7\textwidth]{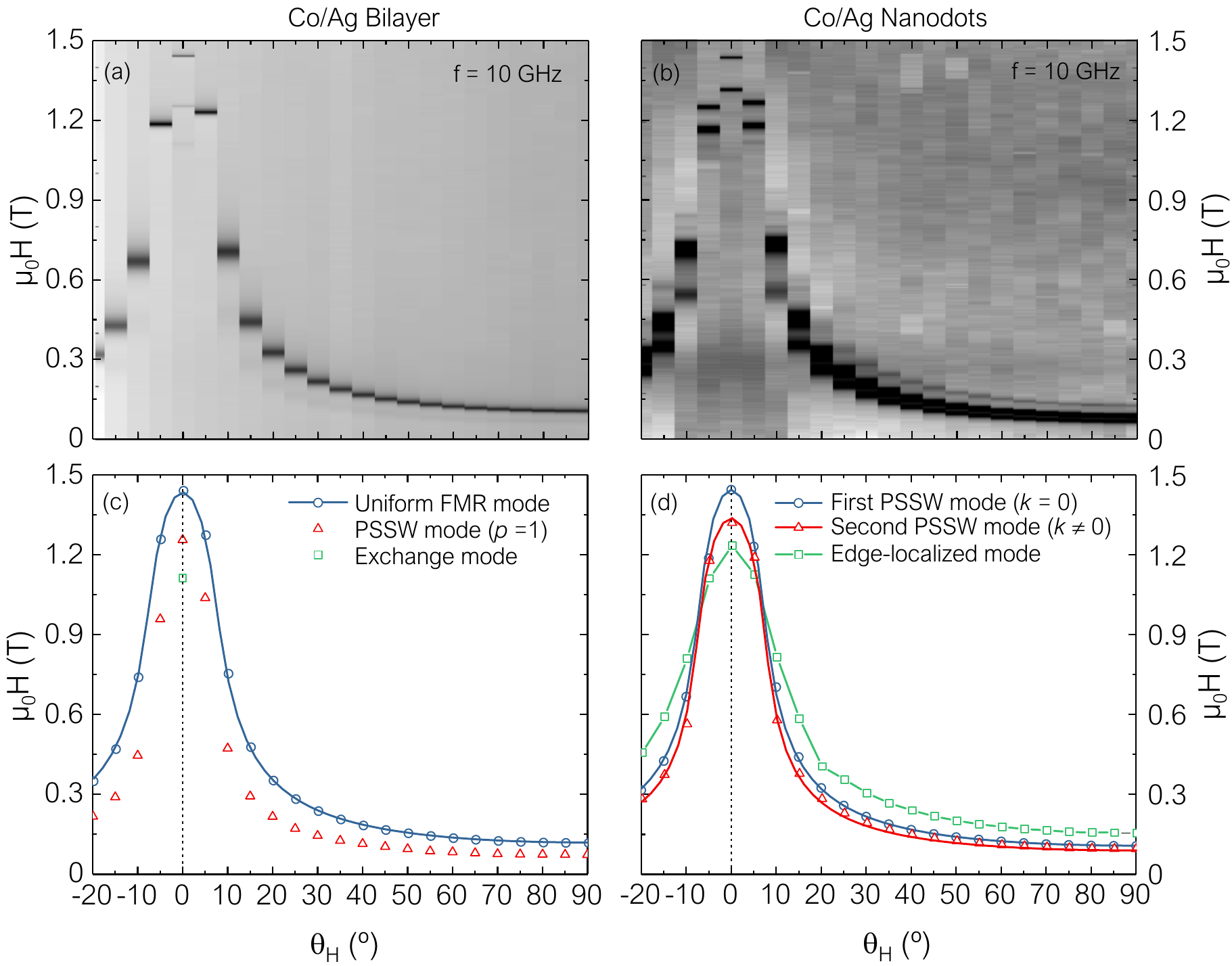}
		\caption{Polar angular dependence of the VNA-FMR signal of the Co/Ag bilayer (a) and nanodots (b) measured at $f$ = \SI{10}{\GHz}. The extracted resonance modes and corresponding fits to Eq.~(\ref{eq:P-FMR}) are depicted in (c) and (d), respectively.}
		\label{fig:Polar-FMR}
	\end{figure*}
	
	Fitting the experimental data to the corresponding resonance equation:
	\begin{widetext}
		\begin{equation}
			\begin{split}
				\bigg( \frac{\omega}{\gamma}\bigg)^2 = &\bigg[\mu_0 H_0 \cos(\varphi-\varphi_H)+\frac{2K_{2\parallel}}{M_s}\cos 2(\varphi-\varphi_u)+\frac{2K_{4\parallel}}{M_s}\cos 4\varphi\bigg] \times \\ 
				&\bigg[\mu_0 H_0 \cos(\varphi-\varphi_H)+ \mu_0 M_\textrm{eff}+\frac{2K_{2\parallel}}{M_s}\cos ^2(\varphi-\varphi_u)+\frac{K_{4\parallel}}{2M_s}(3+\cos 4\varphi)\bigg]
			\end{split}
			\label{eq:A-FMR}
		\end{equation}
	\end{widetext}
	gives the magnetic parameters summarized in Table \ref{tab:AZI}.  
	
	Since the corresponding nanodot data set is not of the same quality as the one of the bilayer and has to be fitted with a dedicated energy model for nanostructured elements, which is not part of our fit software, we decided not to show and not to analyze it. We only note that the spectrum of the nanodots displays the same characteristics as the one of the bilayer, but the data points are generally shifted towards slightly higher resonance fields. 
		
	\begin{table}[t]
		\centering
		\caption{Magnetic parameters obtained by fitting the \mbox{azimuthal} angular dependence of the FMR for the Co/Ag bilayer.}
		\label{tab:AZI}
		\begin{ruledtabular}
			\begin{tabular}{lc}
				Fit Parameter & {Co/Ag Bilayer}\\
				\colrule
				$\mu_0 M_\textrm{eff}$ (T) & 0.975\\
				$\frac{K_{2\parallel}}{M_s}$ (mT) & 2.55\\
				$\frac{K_{4\perp}}{M_s}$ (mT) & -59\\
				$\frac{K_{4\parallel}}{M_s}$ (mT) & 0\\
			\end{tabular}
		\end{ruledtabular}
	\end{table}

	\subsection{Polar Angle-Dependent VNA-FMR}
	
	The polar angular dependence of the FMR was measured at a fixed frequency of $f$ = \SI{10}{\GHz} in steps of \ang{5} for \ang{-20} $\leq \theta_H \leq$ \ang{90}, where $\theta_H$ = \ang{90} and $\theta_H$ = \ang{0} correspond to the previously discussed in-plane and out-of-plane configurations, respectively. The corresponding $H(\theta_H)$ absorption plots of the Co/Ag bilayer and the nanodots are shown in Figs.~\ref{fig:Polar-FMR}(a) and \ref{fig:Polar-FMR}(b), respectively. 
	
	For the bilayer sample, we can clearly identify two resonance modes across the entire range of $\theta_H$: a low field excitation with weak intensity (PSSW mode) as well as a higher field excitation with strong intensity (uniform mode). Only at $\theta_H = 0^\circ$, a third mode (exchange mode) with a resonance field even smaller than that of the low field mode can be observed, which is in line with the number of modes visible in the corresponding out-of-plane VNA-FMR spectrum. The extracted data points and the fit of the uniform mode (first PSSW mode with $k$ = 0) for the bilayer (nanodots) to the resonance equation: \begin{widetext}
		\begin{equation}
			\begin{split}
				\bigg( \frac{\omega}{\gamma}\bigg)^2 = 
				&\bigg[ \mu_0 H_0 \cos(\theta-\theta_H)-\bigg(\mu_0 M_\textrm{eff}+\frac{2K_{2\parallel}}{M_s}-\frac{K_{4\perp}}{M_s}+\frac{K_{4\parallel}}{2M_s}\bigg)\cos 2\theta + \bigg(\frac{K_{4\perp}}{M_s}+\frac{K_{4\parallel}}{2M_s}\bigg)\cos 4\theta \bigg]  \times \\    
				& \bigg[ \mu_0 H_0 \cos(\theta-\theta_H)-\bigg(\mu_0 M_\textrm{eff}+\frac{2K_{2\parallel}}{M_s}-\frac{K_{4\parallel}}{M_s}\bigg)\cos^2\theta 
				+ \bigg(\frac{2K_{4\perp}}{M_s}+\frac{K_{4\parallel}}{M_s}\bigg)\cos^4\theta +\frac{2K_{2\parallel}}{M_s}-\frac{2K_{4\parallel}}{M_s} \bigg]
				\label{eq:P-FMR}
			\end{split}
		\end{equation}
	\end{widetext} 
	are depicted in Figs.~\ref{fig:Polar-FMR}(c) and \ref{fig:Polar-FMR}(d), and the corresponding fit parameters are summarized in Table \ref{tab:POLAR}.  
	
	From the nanodot spectrum in Fig.~\ref{fig:Polar-FMR}(b), a total of three resonance modes can be identified across the entire range of $\theta_H$: two resonances of equally strong intensity (both PSSW modes) as well as a resonance of significantly weaker intensity (edge-localized mode). This is in line with the observations from the in-plane VNA-FMR data, but compared to the out-of-plane geometry, no exchange modes are visible due their increasingly weaker intensity. In contrast to the bilayer sample, there is a mode crossover between $\theta_H = \pm5^\circ$ and $\theta_H = \pm10^\circ$, i.e., in the vicinity of the out-of-plane direction, at which the resonance field of the edge-localized mode drops below the values of the two PSSW modes, while the overall mode intensities remain unchanged. This behavior becomes more obvious when looking at the extracted data points as well as the fits in Fig.~\ref{fig:Polar-FMR}(d).  
	
	\begin{figure*}[ht]
		\centering
		\includegraphics[width=0.9\textwidth]{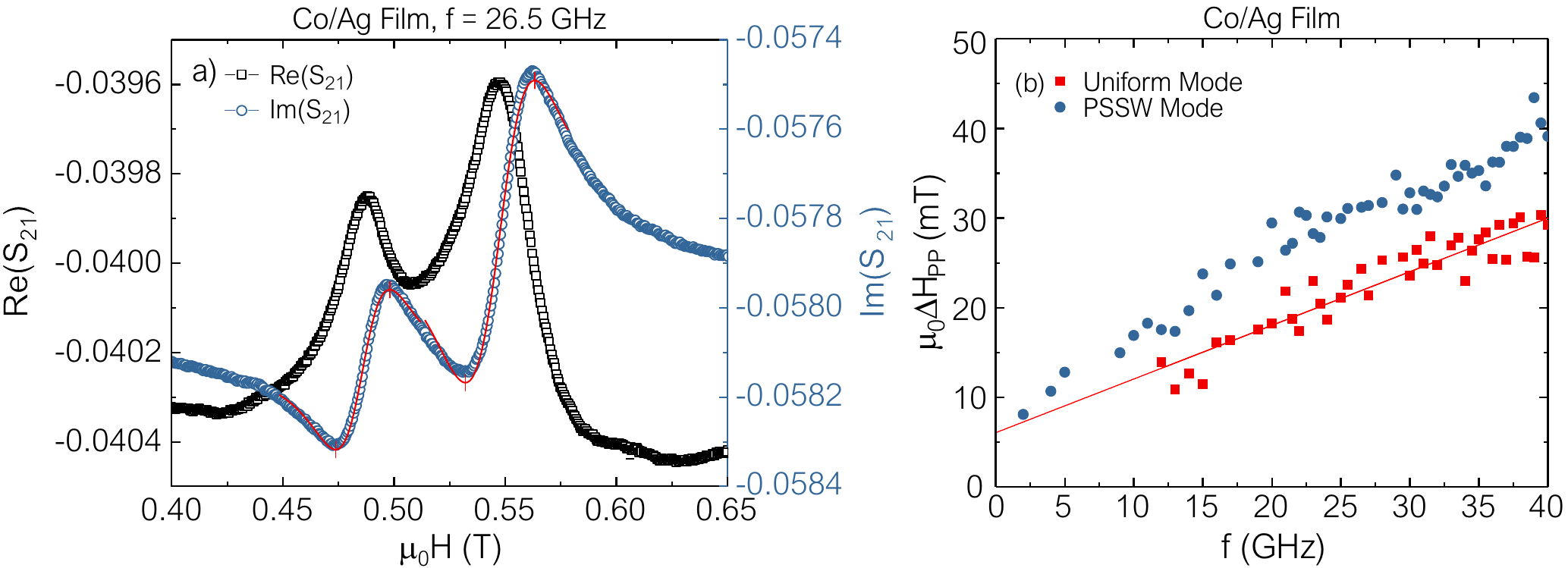}
		\caption{(a) Example of an in-plane easy-axis VNA-FMR spectrum of the Co/Ag bilayer measured at$\ f$ = \SI{26.5}{\GHz} and (b) corresponding peak-to-peak linewidth $\mu_0 \Delta H_\textrm{pp}$, obtained from fitting Im($S_{21}$), as a function of the resonance frequency $f$.}
		\label{fig:Damping}
	\end{figure*}
	
	\begin{table}
		\centering
		\caption{Magnetic parameters obtained by fitting the polar angular dependence of FMR.}
		\label{tab:POLAR}
		\begin{ruledtabular}
			\begin{tabular}{lcc}
				Parameter & {Bilayer} & {Nanodots}\\
				\colrule
				$g$-factor & 2.057 & 2.057  \\
				$\mu_0 M_\textrm{eff}$ (T) & 0.975 & 0.979 \\
				$\frac{K_{2\parallel}}{M_s}$ (mT)& 2.4 & 10.4 \\
				$\frac{K_{4\perp}}{M_s}$ (mT)& 0 & 0 \\
				$\frac{K_{4\parallel}}{M_s}$ (mT) & 0 & 0
			\end{tabular}
		\end{ruledtabular}
	\end{table}

	\subsection{Linewidth and Damping}
	
	Apart from having extracted the resonance frequencies/fields by fitting the experimental data to a complex Lorentzian, the corresponding linewidths have been evaluated as well, from which information about the underlying damping processes in the samples can be obtained. An example of a background-subtracted, in-plane easy-axis VNA-FMR spectrum of the Co/Ag bilayer measured at $f$ = \SI{26.5}{\GHz} is shown in Fig.~\ref{fig:Damping}(a). The corresponding peak-to-peak linewidth $\mu_0 \Delta H_\textrm{pp}$ of Im($S_{21}$) is depicted Fig.~\ref{fig:Damping}(b) for both modes. Thanks to the broadband character of the VNA-FMR measurements, we can immediately see that the main contribution to damping comes in the form of Gilbert-type damping, which manifests itself by a linear dependence between linewidth and frequency. The exact relation between $\mu_0 \Delta H_\textrm{pp}$ and $f$ is given by:
	\begin{equation}
		\mu_0 \Delta H_\textrm{pp} = \frac{4\pi}{\sqrt{3}\gamma}\alpha f + \mu_0 \Delta H_0 ,
		\label{eq:Damping}
	\end{equation}
	where $\alpha$ is the dimensionless (Gilbert) damping parameter and $\mu_0 \Delta H_0$ is an inhomogeneous linewidth broadening, corresponding to the intercept of the fitted linewidth with the $y$-axis at zero frequency. By fitting the linewidth of the uniform mode to Eq.~(\ref{eq:Damping}), we obtain $\alpha = 0.0148$ and $\mu_0 \Delta H_0$ = \SI{6}{mT}. While the linewidth of the PSSW mode with $p = 1$ is larger than the one of the uniform mode, its slope and, thus, its damping parameter is essentially the same. For the nanodot array, a slightly higher value for $\alpha$ as well as a smaller value for $\mu_0 \Delta H_0$ are obtained. In the out-of-plane geometry, we observe the same linear $\Delta H_\textrm{pp}(f)$ dependence, but with both smaller values for $\alpha$ and $\mu_0 \Delta H_0$. Table \ref{tab:Linewidth} summarizes the parameters obtained by fitting the linewidth data for both geometries to Eq.~\ref{eq:Damping}.
	
	\begin{table}[b]
		\centering
		\caption{Damping parameter $\alpha$ and inhomogeneous linewidth broadening $\mu_0 \Delta H_0$ obtained by fitting the both in-plane and out-of-plane VNA-FMR data of the bilayer (uniform mode) and the nanodots (PSSW mode with $p = 1$) to Eq.~(\ref{eq:Damping}).}
		\label{tab:Linewidth}
		\begin{ruledtabular}
			\begin{tabular}{lcccc}
				\multicolumn{1}{c}{} & \multicolumn{2}{c}{Bilayer}& \multicolumn{2}{c}{Nanodots} \\
				\cmidrule{2-3} \cmidrule{4-5} 
				Parameter & IP & OOP & IP & OOP\\		 
				\cmidrule{1-1} \cmidrule{2-3} \cmidrule{4-5}
				$\alpha$ & 0.0148 & 0.0101 & 0.0160 & 0.0057\\
				$\mu_0 \Delta H_0$ (mT) & 6 & 0.7 & 0.6 & 0\\		
			\end{tabular}
		\end{ruledtabular}
	\end{table} 
	\section{Conclusions}
	
	The use of VNA-FMR can be seen to be an extremely powerful tool in the study of the dynamic magnetic properties of magnetic nanostructures and is very sensitive to the generation of standing spin-wave modes. Such studies allow an in-depth characterization of the properties of these systems, however the analysis can be very complex if we wish to make a full interpretation of the experimental data.
	
	In this paper, we have performed detailed VNA-FMR measurements at room temperature in the frequency range up to \SI{40}{\GHz}. The data is quite rich and has been interpreted using the consideration of the boundary conditions of the magnetic nanodots in three dimensions to identify the natural spin-wave excitation modes of the system. These latter have been determined from a consideration of micromagnetic simulations, which have provided us with frequency-dependent modal patterns for the spin-wave excitations in the nanodot system. Once these are taken into account, we have been able to construct a simple model for the standing spin-wave vectors, from which we calculate the spin-wave mode frequencies using the standard theory for spin-wave excitations in ferromagnetic solids, and thus generate the expected spin-wave spectrum of the system. 
	
	We have further used a reference thin-film structure to fit the VNA-FMR data providing us with the magnetic material constants for the Co/Ag nanostructures. From this, we find an excellent agreement for the calculated and experimental values for the frequencies of the fundamental (uniform-like) mode, at around \SI{17.5}{\GHz}. The existence of an edge-localized mode in the experiment has been confirmed and fits very well with theory and micromagnetic simulations, having the form of a flapping mode at the extrema of the nanodot in one of the in-plane directions. Its frequency is below the fundamental mode's frequency and has been shown to be a consequence of the imaginary wave vector for such localized SSW modes. Higher order standing spin-wave modes can be generated from the theory, which allows us to find a probable mode number for the second bulk SSW that lies at frequencies above the fundamental mode. While the assignment of this mode is not entirely unambiguous, both in the framework of the model and the considerations for the spin-wave modes accessible therein, we provide a satisfactory explanation of all the experimental observations. We have illustrated a consistent approach to the full interpretation of experimental measurements, where there is a good degree of confidence and agreement between theory and experiment. 
	
	We are well aware that the combination of theoretical modeling and simulation can be complex and laborious. However, we have provided a consistent model for our nanodot system which is physically feasible and realistic. It is the nature of these excitations that they are inherently complex and we must accept the plausible nature of the agreement between theory and experiment. We furthermore note that, as mentioned previously, while the analytical model used in this paper is an approximation, it provides a “best case” approach for calculating the expected $f$($H$) characteristics in a fully analytical form. Indeed, we cannot expect to obtain an exact solution due to the nature of the excitations over the length scales of the nanodot for such nanostructures based on plane-wave solutions for the spin wave vectors.

	\appendix	
	\section{Micromagnetic Simulation of the High-Frequency Magnetization Dynamics}
	Our three-dimensional simulation study of the magnetic properties of the Co/Ag nanodots is based on a combination of advanced finite-element micromagnetic algorithms that allow us to analyze both the static structure of the magnetization and its high-frequency dynamics in an external rf field. Owing to the large spacing of \SI{200}{\nm}---twice the sample radius---between the disks in the array, we can safely assume that inter-particle magnetostatic interactions are negligible. We thus consider only a single, cylindrical Co sample representing the behavior of the ensemble of disks studied in the experiment. Our model system is a Co disk with \SI{100}{\nm} radius and \SI{30}{\nm} thickness. The numerical model of this sample is discretized into 45874 irregularly shaped tetrahedrons, whose edge length remains below \SI{4}{\nm}, thereby ensuring cell sizes smaller than the material's exchange length $l=\sqrt{2A/\mu_0M_s^2}=4.6\,\si{\nano\meter}$. To model the material properties, we assume a ferromagnetic exchange constant $A=\SI{1.0e-11}{\joule\per\meter}$, a saturation magnetization $\mu_0M_s=\SI{1.096}{\tesla}$, and a weak uniaxial in-plane magnetocrystalline anisotropy $K_u=\SI{8.024}{\kilo\joule\per\meter\cubed}$.
	
	The preliminary step of the method consists in the computation of the magnetic ground state (equilibrium) configuration $\bm{M}_0(\bm{x})$ in the disk-shaped sample for a given constant external field $\bm H_0$, which in our case is oriented in-plane in $x$ direction. We perform this calculation, which is a standard task of any modern micromagnetic simulation software, with our custom-developed, GPU-accelerated finite-element micromagnetic software package \textsc{tetmag} \cite{hertel_large-scale_2019}. The micromagnetic code solves the Landau-Lifshitz-Gilbert (LLG) equation \cite{gilbert_phenomenological_2004}
	\begin{widetext}
		\be \label{gilbert}
		(1+\alpha^2)\frac{d\bm{M}}{dt}=-\gamma\left(\bm{M}\times\bm{H}_\text{eff}\right)-\frac{\alpha}{M_s} \left[ \bm{M}\times\left(\bm{M}\times\bm{H}_\text{eff}\right) \right],
		\ee
	\end{widetext}
	which, starting from an initial magnetic configuration $\bm{M}(\bm{x},t=0)$, is integrated in time until the systems reaches an equilibrium state $\bm M_0(\bm x)$ ($\gamma$ is the absolute value of the gyromagnetic ratio, $\alpha$ is the Gilbert damping constant, and $\bm{H}_\text{eff}$ is the micromagnetic effective field accounting for the energy terms). 
	We remark that, in our computation, the effective field $\bm H_\mathrm{eff}$ in Eq.~\eqref{gilbert} does not include any time-dependent externally applied fields, but only the time-invariant part $\bm H_0$. Thus, \textsc{tetmag} is used to calculate the static magnetic structure $\bm M_0(\bm x)$, whereas the magnetization dynamics driven by an external rf field $\delta\bm H(t)$ superimposed to $\bm H_0$ is treated with a dedicated software that we have developed to specifically address such situations. The principal ingredients of this dynamic code are briefly described in the following sections.
	
	Once the static magnetization structure $\bm{M}_0(\bm{x})$ is obtained, we can numerically probe its response to an external oscillating field $\delta\bm{H}(t)=\delta\hat{\bm{H}} \exp(i\omega t)$, which in our case has an amplitude $\left|\mu_0\delta\hat{\bm{H}}\right|=\SI{0.5}{\milli\tesla}$
	and is oriented in-plane along the $y$ direction, normal to $\bm{H}_0$. 
	To this end, the magnetization dynamics is decomposed into a static and a fluctuating component:
	\be\label{linear}
	\bm{m}(\bm{x},t)=\bm{m}_0(\bm{x})+\delta{\bm{\hat{m}}(\bm{x})}\exp(i\omega t),\\
	\ee 
	where $\left|\delta\hat{\bm{m}}\right|\ll 1$ and $\bm{m} = \bm{M}/M_s$ ($\bm m$ is a unit-vector field). A corresponding linear approximation is also done for the effective micromagnetic fields. In our dynamic simulations, we set the value of the Gilbert damping constant to $\alpha=0.01$.
	The ansatz expressed by Eq.~(\ref{linear}) allows us to linearize the LLG equation by retaining only first-order terms and to reformulate the problem of finding small-angle harmonic oscillations $\delta\bm{m}$ of the magnetization around its equilibrium position $\bm{m}_0$ in the form of a linear system in the frequency domain \cite{vukadinovic_ferromagnetic_2001,daquino_novel_2009}: 
	\be\label{linear_eq}
	{\cal L}\delta\hat{\bm{m}} = {\cal P} \delta\hat{\bm{H}},
	\ee
	where ${\cal L}={\cal L}[\bm{m}_0, \omega, \alpha]$ is a linear operator containing the system's dynamic interactions in the form of micromagnetic effective fields and ${\cal P}$ is a constant projection operator \cite{daquino_novel_2009}. The operator ${\cal L}$ 
	has properties similar to those of a ``dynamic matrix'' as it is commonly used in comparable algorithms \cite{Born1954, vukadinovic_ferromagnetic_2001}. 
	When discretized on a computational grid with $N$ cells, the operator $\cal L$ becomes a fully-populated matrix with $\mathcal{O}(N^2)$ dimension ($N$ is the number of grid cells), which rapidly saturates the available computer memory as $N$ becomes moderately large and makes simulation of large magnetic systems unfeasible.
	Nevertheless, by using an appropriate operator formalism developed by d'Aquino {\em et al.}~\cite{daquino_computation_2008,daquino_novel_2009}, one can solve Eq.~\ref{linear_eq} without storing the dense matrix associated with $\mathcal{L}$, resulting in a matrix-free formulation. Moreover, this approach allows to exploit 
	several modern features such as GPU-acceleration of the magnetostatic field calculation and implementation of a ${\cal H}2$-type hierarchical matrix method to treat large-scale problems \cite{hackbusch_hierarchical_2015} (featuring $\mathcal{O}(N)$ storage and computational cost) that would otherwise entail prohibitively large memory requirements and computation times. We will refer to this novel method as frequency domain, matrix-free micromagnetic linear response solver (MF-\textmu LRS).
	
\begin{figure}
		\centering
		\includegraphics[width=0.48\textwidth]{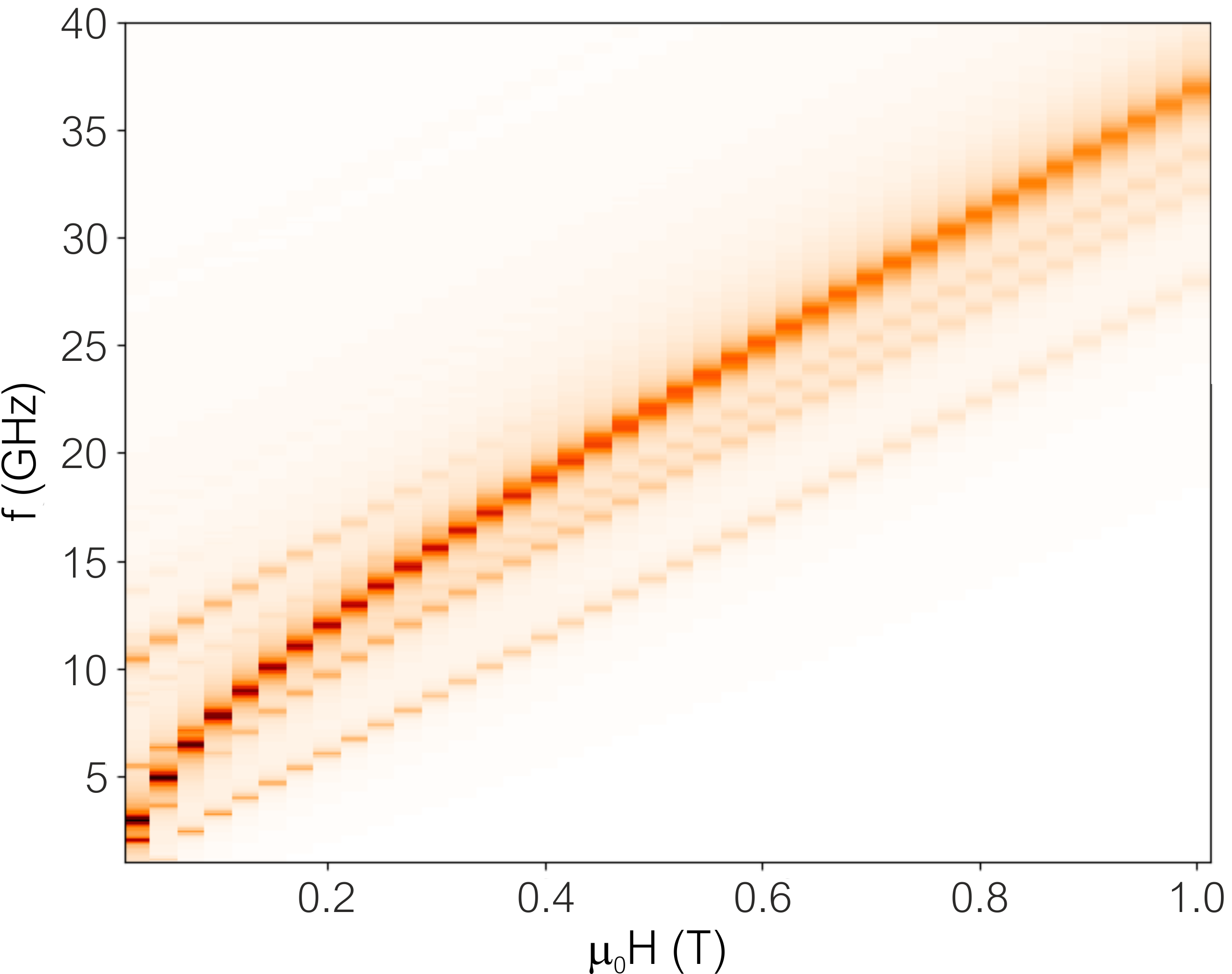}
		\caption{In-plane $f(H)$ plot for the simulation of the nanodot structure using the material parameters referred to in the text.}
		\label{fig:Simulations}
	\end{figure}

	In Fig.~\ref{fig:Simulations}, we show the results for the in-plane $f(H)$ plot for the resonance modes in the nanodot structure. A detailed discussion of the simulations and the interpretation of the resonance experiment is given in Section \label{Nanodot arrays}.
	
	Due to space limitations, and in order to preserve the focus of the article on the physical properties of the system studied here, we cannot explain more numerical or mathematical details of our frequency domain MF-\textmu LRS algorithm here. A more exhaustive description can be found in Ref.~\cite{d’Aquino2023}. 
	
	Even for relatively large micromagnetic problems, the numerical solution of Eq.~(\ref{linear_eq}) is sufficiently fast that we can obtain a quasi-continuous set of solutions of the dynamic magnetization $\delta\bm{m}_{\omega_i}(\bm{x})$ for a large set of frequencies $\omega_i$, each yielding a different operator ${\cal L}^\omega_i$. We typically ``sweep'' the frequency $\omega_i$ in steps of $\Delta\omega=\SI{100}{\MHz}$ within a range of \SIrange{0.1}{20}{\GHz} for each static field value $\bm{H}_0$. In all cases, the amplitude of the oscillating external field, $|\mu_0\delta\hat{\bm{H}}|$, remains constant at a value of \SI{0.5}{\milli\tesla}, oriented parallel to the $y$-axis. 
	\begin{figure*}[ht]
		\centering
		\includegraphics[width=0.75\textwidth]{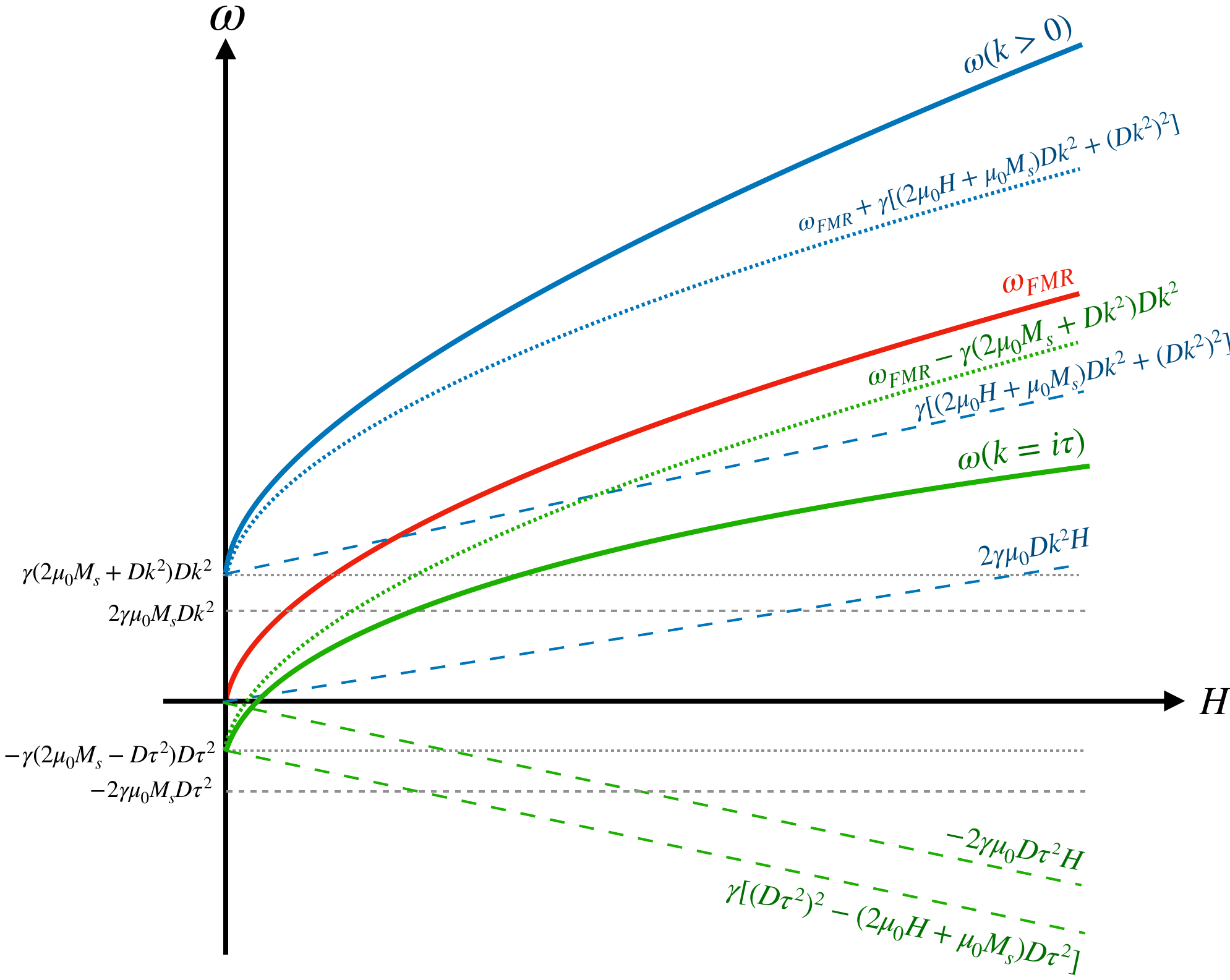}
		\caption{Variation of the angular frequency $\omega$ as a function of the applied magnetic field $H$.}
		\label{fig:Theor}
	\end{figure*}
	An alternative to the aforementioned frequency-domain MF-\textmu LRS approach lies in performing simulations of the LLG Eq.~\eqref{gilbert} driven by suitable field pulses and post-processing the time-domain magnetization response via Fourier transforms. More specifically, the simulation would consist in shifting the static magnetic structure $\bm{M}_0$ slightly out of equilibrium by means of a small external perturbation such as, e.g., a picosecond field pulse and letting the magnetization relax back towards equilibrium for a sufficiently long time~\cite{Yan2007, mcmichael_magnetic_2005}. This can be done by using traditional micromagnetic software such as our \textsc{tetmag} code. The time-domain dynamics of the magnetization can then be Fourier-analyzed, yielding the power spectrum of the system. The results of this so-called ``ringdown'' method \cite{baker_proposal_2017} are equivalent to those of our frequency domain MF-\textmu LRS, but the former is numerically much more time-consuming and bears the risk of accumulating numerical inaccuracies of the time integration of the LLG equation over a long time span (several \si{\ns}) at low damping. We have verified on a few examples that the ringdown method and our dynamic-matrix method yield identical resonance frequencies and mode profiles.

	\section{Evaluation of Standing Spin-Wave Modes and Surface Modes}
	
	\begin{figure*}[ht]
		\centering
		\includegraphics[width=0.90\textwidth]{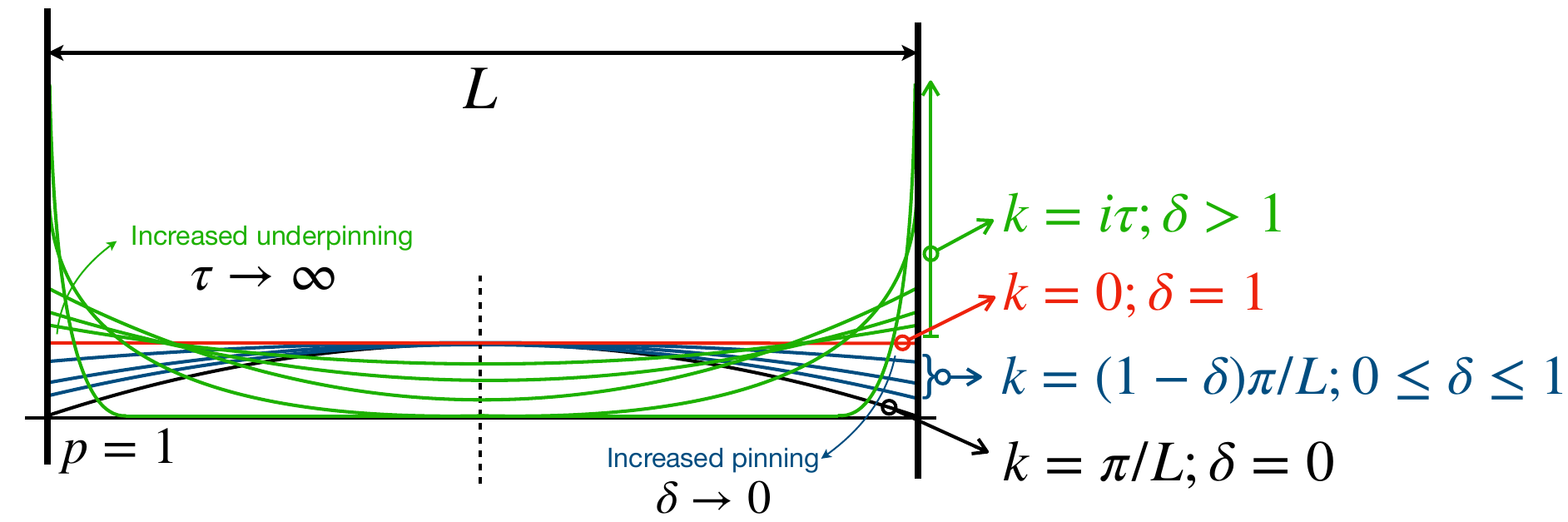}
		\caption{Schematic representation of the 1D $p=1$ modes. Perfect freedom corresponds to the uniform resonance mode with $k=0$ and $\delta =1$ (red). For increased pinning, $k>0$ and $\delta \rightarrow 1$ the mode forms a volume standing spin wave (VSSW) in blue and at the limit $\delta = 0$, we have the case for perfect pinning (black). Underpinning is illustrated in green, where the imaginary wave vector, $k = i\tau$, increases and leads to greater degree of localization of the mode.}
		\label{fig:p=1modes}
	\end{figure*}
	
	It is instructive to outline the various possible modes that can arise in ferromagnetic resonance experiments. The nature of these modes will depend on a number of factors including the material parameters of the sample under investigation. Critical to such considerations are the boundary conditions for samples with reduced dimensions, such as this films and nanostructures, typically of the order of tens of nanometers in size. These boundary conditions have been discussed in Section \ref{Nanodot arrays}. In this appendix, we aim to outline the analysis of the resonance equation to illustrate the various contributions and how we can use this analysis to show the form and behavior of the uniform mode of resonance (or the pure FMR mode), bulk standing spin wave modes and localized and surface resonance modes. 
	
	In its simplest form, the resonance equation can be expressed as, see Eq.~(\ref{eq:fmr1}):
	\begin{equation}
		\left(\frac{\omega }{\mu_0 \gamma}\right)^2 = (H + M_\textrm{s} + Dk^2)(H + Dk^2).
		\label{eq:fmr2}
	\end{equation}
	
	Let us consider the various forms of solution for this expression. In the simplest case, for the uniform resonance mode, the wave vector $k=0$. Strictly speaking, this is ferromagnetic resonance, in which all spins in the magnetic system precess in phase with the exact same amplitude of precession. This simple case can be represented as a single magnetization vector in the classical sense. In this case, the above resonance equation is simplified to:
	\begin{equation}
		\left(\frac{\omega_\text{FMR} }{\gamma}\right)^2 = \mu_0 H (\mu_0 H + \mu_0 M_\textrm{s})
		\label{eq:fmr3}
	\end{equation}
	or 
	\begin{equation}
		\omega_\text{FMR} =\gamma \mu_0 \sqrt{H ( H + M_\textrm{s})} = \gamma \sqrt{\mu H B}.
		\label{eq:fmr3b}
	\end{equation}
	
	The variation of the resonance frequency ($f = \omega/2\pi$) is illustrated in Fig.~\ref{fig:Theor} with the red line. This curve passes through the origin.

	Since the case with non-zero wave vectors is the most general form, we can express Eq.~(\ref{eq:fmr2}) in terms of the FMR response plus additional terms due to the existence of spin-wave modes. Thus we write:
	\begin{equation}
		\left(\frac{\omega }{\mu_0\gamma}\right)^2 = \left(\frac{\omega_\text{FMR} }{\mu_0 \gamma}\right)^2 +Dk^2 (2 H + M_\textrm{s} ) + (Dk^2)^2.
		\label{eq:fmr4}
	\end{equation}
	
	This can be considered as a shift of the resonance line with respect to the FMR line, as illustrated in Fig.~\ref{fig:Theor} with the blue line. The various contributions to the general expression are also illustrated in the dashed and dotted blue lines. We note that this equation is quadratic in the spin-wave term $Dk^2$. The general solution for this can be expressed in the usual manner as:
	\begin{widetext}
		\begin{equation}
			Dk^2 = -\frac{(2 H +  M_\textrm{s} )}{2} \pm \frac{\sqrt{(2 H +  M_\textrm{s} )^2 - 4 (\omega^2 - \omega_\text{FMR}^2)/(\mu_0\gamma)^2}}{2}.
			\label{eq:SWsoln}
		\end{equation}
	\end{widetext}
	
	These solutions can be separated into real solutions, which correspond to \textit{bulk} or \textit{volume modes}, where the spin precession occurs through the body of the magnetic nanostructure, and into \textit{localized modes}, where the spin precession is strongest at the boundaries of the magnetic structure and is weak or zero in its center. For the former, it is clear that we must have a positive ($+$) solution in which the second term of Eq.~(\ref{eq:SWsoln}) is greater than the first, we can denote these wave vectors as $k = k_{\alpha}$. For localized modes, there are two forms than can be distinguished: (a) those for which we have a negative ($-$) solution to Eq.~(\ref{eq:SWsoln}), and (b) those for which we have a positive ($+$) solution to Eq.~(\ref{eq:SWsoln}), where the second term of Eq.~(\ref{eq:SWsoln}) is smaller than the first. In these two cases for localized excitations, the wave vector must be imaginary, and we can write $k =  k_{\beta, a} = i\tau_{a}$ for (a) and $k =  k_{\beta, b} = i\tau_{b}$ for (b). For further discussion into the nature of physically real solutions see Ref.~\cite{Maksymowicz1986}. For our purposes, we can simply take the fact that we can expect, for certain boundary conditions, the existence of edge or surface localized modes, for which we can express the wave vector in the form $k = i\tau$. Adopting this notation allows us then to write the corresponding resonance equation in the form:
	\begin{equation}
		\left(\frac{\omega }{\mu_0 \gamma}\right)^2 = \left(\frac{\omega_\text{FMR} }{ \mu_0 \gamma}\right)^2 -D\tau^2 (2 H + M_\textrm{s} ) + (D\tau^2)^2.
		\label{eq:fmr5}
	\end{equation}
	
	Such solutions can again be seen to cause a shift in the resonance frequencies with respect to the uniform mode, and are illustrated in Fig.~\ref{fig:Theor} as the green solid line. The various contributions to the general expression are also illustrated in the various dashed and dotted green lines. We further note that while the green curve in the figure appears negative, this is only a consequence of the mathematical solution and does not correspond to a physically real solution in the negative frequency region. 
	
	Comparing the solutions illustrated by the solid lines in Fig.~\ref{fig:Theor}, we see that the volume standing spin wave (VSSW) modes [$\omega(k>0)$] are up-shifted with respect to the uniform mode ($\omega_\text{FMR}$), while the localized modes [$\omega(k = i\tau)$] are frequency down-shifted with respect to the uniform mode. While we only illustrate one of each type of VSSW and a localized mode, in reality there can be several additional modes depending on the mode numbers and geometries, as discussed in Section \ref{Nanodot_array}. In Fig.~\ref{fig:p=1modes}, we show the $p=1$ mode profiles as a function of the pinning conditions. Here we show the uniform ($k=0$) mode in red, underpinning, with increasing pinning at the boundaries, $\delta$, leads to VSSW modes, up to the limit of $\delta = 0$, giving the case of perfect pinning. Underpinning (green lines) leads to greater localization of the mode at the boundaries.

	\section{Evaluation of the Lowest Lying Standing Spin-Wave Modes}
	
	In Section \ref{Nanodot_array}, we define the expected form of the wave vectors for our nanodot structures, which are based on the form of the simulated excitations for a nanodot. This wave vector is expressed as:
	\begin{equation}
		\begin{split}
			k^2_{pqr} &= \bigl[p ^2 + (q - 1)^2 \bigr] \left(\frac{ \pi}{d}\right)^2 +(r -1)^2\left(\frac{\pi}{L}\right)^2\\ &= \bigl[p ^2 + (q - 1)^2 \bigr] n_2 +(r -1)^2 n_1.
			\label{eq:WVmixedboundAp}
		\end{split}	
	\end{equation}
	
	Here we have been explicit in our choice of the form of the wave vector so as to produce modal patterns close to that of the simulation. Once we have made this choice, it is now only a question of performing the calculations to generate the spectrum of excitation modes. Here we will limit ourselves to the lowest lying modes, since these will be closest to the fundamental mode and in the range of measured values. Despite this, we should take care to generate enough modes to make sure that we allow for the expected degeneracy between modes. For illustrative purposes, we will only consider modes with $r=0$ and $r=1$. In the following table, we evaluate the values of the wave vectors generated from Eq.~(\ref{eq:WVmixedboundAp}).
	
	We note that even for this limited set of wave vectors, there is a high degree of mode degeneracy, i.e., modes with the same value of $k^2_{pqr}$. For example, we have $k^2_{001} =k^2_{111} = k^2_{021}$, etc. At the end of the table, we include some selected values of $pqr$ for which the $k^2$ values are relatively low. The 000 mode is actually a high order mode in the sense that it has a high $k^2$ value, while for modes with $r=1$, we note that the  $k^2$ values are relatively low. This is a consequence of the pinning parameter choice and the resulting form of the wave vector, Eq.~(\ref{eq:WVmixedboundAp}).
	
	From Table \ref{tab:Mode numbers} we can evaluate the expected spectrum for the nanodots. This will consist of the sequence of lines for which $Dk^2 = 0, Dn_2, 2Dn_2, 4Dn_2, 5Dn_2, 9Dn_2, ...$ etc. This can be used to determine the frequency spectrum using the resonance equation, see for example Eq.~(\ref{eq:fmr4}). 
	
	\begin{table}[t]
		\centering
		\caption{Lowest lying spin-wave modes for the model given in Eq.~(\ref{eq:WVmixedboundAp}).}
		\label{tab:Mode numbers}
		\begin{ruledtabular}
			\begin{tabular}{ccc}
				$p$, $q$, $r$ & $k^2_{pqr} $& {$k^2_{pqr}$ ($\times 10^{15}$\:m$^{-2}$)} \\
				\colrule
				0, 0, 0 & $n_2 + n_1$ & 4.20 \\	
				&  &   \\	 
				0, 0, 1 & $n_2 $ & 0.247 \\
				0, 1, 0 & $n_1$ & 3.95 \\
				1, 0, 0 & $n_2 + n_1$ & 4.20 \\
				&  &   \\
				0, 1, 1 & $0$ & 0 \\
				1, 0, 1 & $2n_2$ & 0.494\\
				1, 1, 0 & $n_2 + n_1$ & 4.20 \\	
				&  &   \\
				1, 1, 1 & $n_2 $ & 0.247 \\	
				&  &   \\ 
				0, 0, 2 &  $n_2 + n_1 $ & 4.20 \\
				0, 2, 0 &  $n_2 + n_1$ & 4.20 \\
				2, 0, 0 &  $5n_2 + n_1$ & 5.185 \\
				&  &   \\
				0, 1, 2 &  $n_1 $ & 3.95 \\
				1, 0, 2 &  $2n_2 + n_1$ & 4.44 \\
				1, 2, 0 &  $2n_2 + n_1$ & 4.44 \\
				0, 2, 1 &  $n_2  $ & 0.247 \\
				2, 0, 1 &  $5n_2 $ & 1.235 \\
				2, 1, 0 &  $4n_2 + n_1$ & 4.94 \\
				&  &   \\
				1, 1, 2 &  $n_2 + n_1 $ & 4.20 \\
				1, 2, 1 &  $2n_2$ &0.494 \\
				2, 1, 1 &  $4n_2$ &0.988 \\
				&  &   \\
				0, 2, 2 &  $n_2 + n_1 $ & 4.20 \\
				2, 0, 2 &  $5n_2 + n_1$ & 5.185 \\
				2, 2, 0 &  $5n_2 + n_1$ & 5.185 \\
				&  &   \\
				1, 2, 2 &  $2n_2 + n_1 $ & 4.44 \\
				2, 1, 2 &  $4n_2 + n_1$ & 4.94 \\
				2, 2, 1 &  $5n_2 $ & 1.235 \\
				&  &   \\
				2, 2, 2 &  $5n_2 + n_1$ & 5.185 \\
				: & :&  :\\
				0, 3, 1 &  $4n_2 $ & 0.988 \\
				1, 3, 1 &  $5n_2$ & 1.235 \\
				0, 4, 1 &  $9n_2  $ & 2.223 \\
				3, 1, 1 &  $9n_2 $ & 2.223 \\
				3, 0, 1 &  $10n_2$ & 2.47\\
				:& :& :\\
			\end{tabular}
		\end{ruledtabular}
	\end{table}

\begin{acknowledgments}
D.S.S.~and D.M.~acknowledge financial support from the Institut de Physique of CNRS for experimental equipment and a postdoc position, respectively. This work was funded by the French National Research Agency (ANR) through the Programme d'Investissement d'Avenir under contract ANR-11-LABX-0058\_NIE and ANR-17-EURE-0024 within the Investissement d’Avenir program ANR-10-IDEX-0002-02. R.H.~and R.C.~acknowledge the High Performance Computing center of the University of Strasbourg for supporting this work by providing access to computing resources. This project has been supported by the COMET K1 centre ASSIC Austrian Smart Systems Integration 100 Research Center. The COMET --- Competence Centers for Excellent Technologies --- Program is supported by BMVIT, 101 BMDW and the federal provinces of Carinthia and Styria.
\end{acknowledgments}

\pagebreak
\bibliography{Nanodots_PhysRevAppl.bib}

\end{document}